\documentclass[
preprint,
 amsmath,amssymb,
 aps, physrev,
]{revtex4-2}

\usepackage{graphicx}
\usepackage{dcolumn}
\usepackage{bm}
\usepackage{hyperref}

\usepackage{caption}
\usepackage{xcolor}
\usepackage{diagbox}
\usepackage{booktabs}
\usepackage{adjustbox}
\usepackage[mathlines]{lineno}
\usepackage{amsthm}
\newtheorem{observation}{Observation}[section]

\newtheorem{lemma}{Lemma}[section]
\newtheorem{remark}{Remark}[section]
\usepackage[linesnumbered,ruled,vlined]{algorithm2e}
\usepackage{braket}
\usepackage{multirow}
\usepackage{subcaption}
\usepackage{tikz}
\newcommand{\E}{\mathbb{E}}
\newcommand{\G}{\mathbb{G}}
\newcommand{\I}{\mathbb{I}}
\newcommand{\ketbra}[1]{|#1\rangle\langle#1|}

\newcommand{\N}{\mathbb{N}}
\newcommand{\R}{\mathbb{R}}
\newcommand{\Z}{\mathbb{Z}}
\newcommand{\tr}{\text{tr}}
\usepackage{ragged2e}
\SetAlgoCaptionLayout{raggedright}

\begin{document}


\title{\textbf{Scalable bayesian shadow tomography for quantum property estimation with set transformers}}%

\author{Hyunho Cha}
\email{Contact author: ovalavo@snu.ac.kr}
\affiliation{NextQuantum and Department of Electrical and Computer Engineering, Seoul National University, Seoul 08826, Republic of Korea}
\author{Wonjung Kim}
\email{dnjswnd116@snu.ac.kr}
\affiliation{NextQuantum and Department of Electrical and Computer Engineering, Seoul National University, Seoul 08826, Republic of Korea}
\author{Jungwoo Lee}
\email{Corresponding author: junglee@snu.ac.kr}
\affiliation{NextQuantum and Department of Electrical and Computer Engineering, Seoul National University, Seoul 08826, Republic of Korea}

\date{\today}

\begin{abstract}
  A scalable Bayesian machine learning framework is introduced for estimating scalar properties of an unknown quantum state from measurement data, which bypasses full density matrix reconstruction. This work is the first to integrate the classical shadows protocol with a permutation-invariant set transformer architecture, enabling the approach to predict and correct bias in existing estimators to approximate the true Bayesian posterior mean. Measurement outcomes are encoded as fixed-dimensional feature vectors, and the network outputs a residual correction to a baseline estimator. Scalability to large quantum systems is ensured by the polynomial dependence of input size on system size and number of measurements. On Greenberger-Horne-Zeilinger state fidelity and second-order R\'enyi entropy estimation tasks---using random Pauli and random Clifford measurements---this Bayesian estimator always achieves lower mean squared error than classical shadows alone, with more than a 99\% reduction in the few copy regime.
\end{abstract}

\maketitle


\section{Introduction}

Quantum technologies promise advances in communication, computation, and sensing by harnessing phenomena such as superposition and entanglement \citep{montanaro2016quantum, degen2017quantum, cozzolino2019high}. Because quantum protocols rely heavily on precise quantum state preparation, the ability to accurately characterize quantum states formed in the laboratory or on noisy quantum processors is central to the progress of these technologies. The exponential growth of a quantum system's Hilbert space with qubit number renders many existing state tomography protocols infeasible beyond a handful of qubits: reconstructing a full density matrix inevitably requires an exponential amount of computation in the worst case \citep{gross2010quantum, haah2016sample, o2016efficient}.

To address this scalability issue, shadow tomography circumvents full-state reconstruction by focusing on estimating only those scalar properties of a state (mostly linear functions) that are of experimental interest \citep{aaronson2018shadow, huang2020predicting}. By sampling randomized measurements and computing simple linear inversion estimators, \emph{classical shadows} provide accurate estimates of expectation values and certain non-linear functions using few measurements. Nonetheless, these estimators remain non-Bayesian, often exhibiting high variance for a finite number of measurements and lacking a principled way to incorporate prior knowledge about the state.

\begin{figure}[t]
    \centering
    \includegraphics[width=0.9\linewidth]{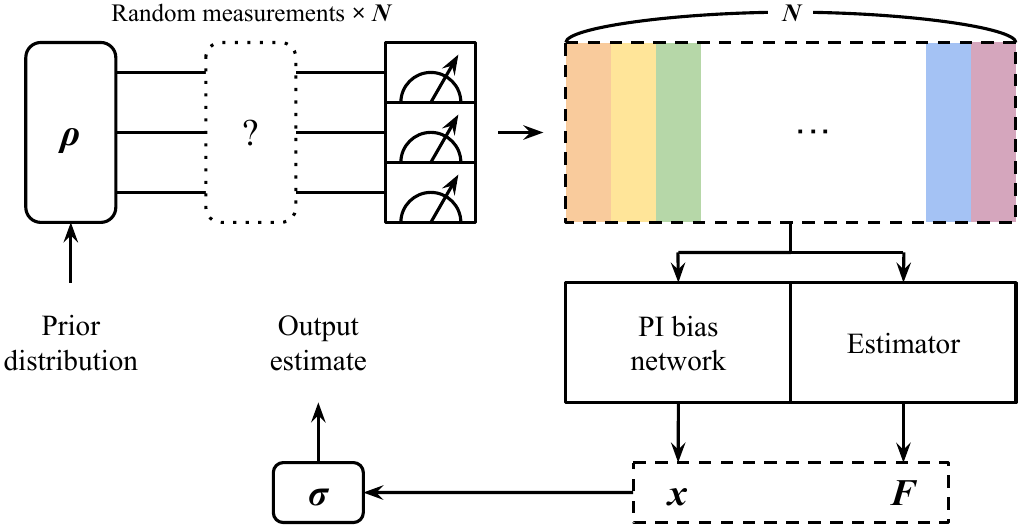}
    \caption{\justifying General framework for residual Bayesian estimation of functions of quantum states. A single experimental realization, comprising \(N\) measurement outcomes, is encoded into \(N\) feature vectors characterizing the outcomes. A PI network is trained to infer a quantity \(x\) that depends on the measurement data and corrects the bias introduced by a baseline non-Bayesian estimator, which outputs a point estimate \(F\).}
    \label{fig:schematic}
\end{figure}

This work investigates Bayesian shadow tomography, combining motivations from Bayesian quantum state estimation with the classical shadows protocol. In shadow tomography, the goal is not to reconstruct the entire quantum state. Instead, it focuses on estimating specific properties or ``shadows'' of the state directly. Our objective is to extend this paradigm to the Bayesian estimation setting. Although Bayesian approaches to quantum tomography are present in the literature, the body of work remains relatively limited \citep{buvzek1998reconstruction, schack2001quantum, granade2016practical, lukens2020practical}. Most existing methods focus on reconstructing the quantum state itself, which inherently requires an exponential amount of data and computation. To the best of our knowledge, we are the first to propose a Bayesian and scalable machine learning approach that directly estimates scalar functions of an unknown quantum state without performing full state reconstruction. This is achieved by efficiently encoding each of the \(N\) measurement outcomes into a fixed-dimensional feature vector, so that the model input grows linearly in \(N\) and only polynomially in the qubit number \(n\)---namely \(O(n)\) for random Pauli measurements and \(O(n^2)\) for random Clifford measurements. This encoding makes it possible to train permutation-invariant (PI) networks that scale to large \(N\) and large \(n\) without explicit reference to the \(2^n\)-dimensional Hilbert space (see FIG.~\ref{fig:schematic}). The use of PI models to account for the unordered nature of random measurement data has proven effective in classification tasks \citep{tang2024ssl4q}. However, our experiments suggest that their direct extension to regression tasks yields limited performance. We suspect this difficulty arises because shadow inversion involves evaluating parity-like functions over measurement outcomes---an extremely rugged mapping for neural networks to capture. Instead, we formulate our estimator via residual learning, so that the network only needs to learn a correction term on the top of a fixed baseline.

Our method employs a non-adaptive measurement strategy, meaning the measurement settings are fixed in advance and not adjusted based on previous measurement outcomes. We focus on the post-processing or final estimation stage of an experiment that has already been conducted, rather than on adaptively collecting data. The classical shadows protocol itself is essentially a form of non-adaptive linear inversion tomography \citep{sugiyama2013precision, guctua2020fast, huang2020predicting}. Moreover, the shadow estimation procedure operates without incorporating any prior information about the state, and individual snapshots of the state can even correspond to non-physical density operators. In other words, classical shadow tomography does not perform a Bayesian update. Nevertheless, any data obtained from measurements can be used to update a prior distribution over the state. Therefore, inspired by the success of classical shadow tomography, we propose to leverage measurement statistics and shadow estimation algorithms within a Bayesian tomography framework.

The remainder of this paper is organized as follows. In Section~\ref{sec:background}, notations and prior Bayesian and ML-based tomography methods are introduced. Section~\ref{sec:problem_statement} states our Bayesian tomography problem. In Section~\ref{sec:methods} we present our PI residual-learning framework, including the encoding schemes for measurement records. Section~\ref{sec:experiment_results} describes experimental results. Finally, Section~\ref{sec:discussion} discusses implications and future directions.

\section{Preliminaries and related works}
\label{sec:background}

\subsection{Notations}

We first introduce notation and relevant background. Let $n$ denote the number of qubits in the system, so that the Hilbert space has dimension $2^n$ (denoted $\mathcal{H}$). The set of bounded linear operators on $\mathcal{H}$ is denoted by $\mathcal{B}(\mathcal{H})$. The set of quantum states (density operators) on $\mathcal{H}$ is denoted by $\mathcal{S}(\mathcal{H})$, defined as all positive semi-definite operators with unit trace:
\[
\mathcal{S}(\mathcal{H}) := \{ \rho \in \mathcal{B}(\mathcal{H}) \mid \rho \succeq 0, \, \tr (\rho) = 1 \}.
\]
The set of quantum observables on \(\mathcal{H}\), denoted \(\mathcal{O}\), consists of Hermitian operators:
\[
\mathcal{O}(\mathcal{H}) := \{ O \in \mathcal{B}(\mathcal{H}) \mid O = O^\dagger \}.
\]
The $2 \times 2$ identity operator on a single qubit is denoted by \(\I\). Quantum basis states are written using the tensor product notation. For example, the \(n\)-qubit computational basis state where all qubits are in state \(\ket{0}\) is denoted by
\[
\lvert 0 \rangle^{\otimes n} := \lvert 0 \rangle \otimes \cdots \otimes \lvert 0 \rangle \in \mathcal{H}.
\]
We use the shorthand \(\ket{\mathbf{b}} \in \mathcal{H}\), where \(\mathbf{b} \in \{0,1\}^n\), to denote the standard computational basis states. For notational convenience, \(\mathbf{b}\) is also interpreted as the corresponding integer in \([0,2^n)\) when the context is clear. For a basic introduction to quantum states, density matrices, and the classical shadows protocol, refer to Appendices~\ref{sec:quantum_state_density_matrix_intro} and \ref{sec:classical_shadow_intro}.

\subsection{Prior works}

Quantum state tomography (QST) aims to reconstruct the quantum state or infer its properties from measurement data. Traditional Bayesian approaches to QST update a prior distribution over the density operator using observed outcomes. Such methods yield principled uncertainty quantification but often suffer from computational cost \citep{blume2010optimal, granade2016practical, lukens2020practical}. A recent effort has mitigated this challenge via neural-network–enhanced Bayesian inference \citep{lohani2023demonstration}, but it remains focused on full state reconstruction.

Parallel to Bayesian methods, non‑Bayesian machine‑learning techniques have been developed to estimate scalar functions of quantum states directly. LLM4QPE \citep{tang2024towards} is an autoregressive model that treats quantum property estimation as a language task to predict various properties. QUADIM \citep{tangquadim} adopts a diffusion-based approach for the same purpose. However, these models assume access to a classical description of the system's evolution and do not enforce permutation invariance of measurement records.

Another line of work employs neural networks to accelerate full-state QST through adaptive measurement strategies or compressed representations \citep{torlai2018neural, quek2021adaptive, lange2023adaptive}. More recently, transformer-based architectures have been used in reconstructing density matrices from measurement data \citep{zhong2022quantum, palmieri2024enhancing, wei2024neural, ma2025tomography}. While these methods achieve remarkable accuracy, they often focus on recovering the density matrix rather than directly targeting scalar quantities. This incurs unnecessary computational overhead when only a few properties are of interest.

This work develops a Bayesian inference framework enhanced by residual learning and implemented with a PI neural network architecture, aiming to estimate scalar functions of an unknown quantum state from classical measurement data. Rather than reconstructing the full density matrix, our method focuses on efficient, scalable estimation of quantities such as fidelity and R\'enyi entanglement entropy. It exploits the unordered nature of non-adaptive measurements to design an architecture that generalizes for various system sizes as well as the numbers of measurements.

\section{Problem setup}
\label{sec:problem_statement}

Consider the following setting for Bayesian quantum tomography. We are given many copies of an unknown $n$-qubit quantum state $\rho$, drawn from some prior distribution (representing our initial knowledge about $\rho$). Our aim is to estimate a scalar function $f(\rho)$---for instance, the expectation value of a particular observable---using data from $N$ projective measurements performed on independent copies of $\rho$.

The feature vector summarizing the outcome of the $k$-th measurement ($1 \le k \le N$) is denoted as $\mathbf{v}^{(k)}$. The structure of $\mathbf{v}^{(k)}$ depends on the measurement scheme; for example, it may encode the measurement basis and the obtained bitstring result. Let $\mathbf{v}^{(1)}, \dots, \mathbf{v}^{(N)} \in \R^d$ denote the sequence of measurement outcomes. It is assumed that there exists an estimator function $g$ which maps the sequence of outcomes to an estimate of the quantity of interest. In other words,
\[
\hat{f}(\rho) = g \left( \mathbf{v}^{(1)}, \ldots , \mathbf{v}^{(N)} \right)
\]
produces an estimate $\hat{f}(\rho)$ of $f(\rho)$ from the data. A simple choice of $g$ in many cases is the sample mean of some function of each outcome (which is the essence of the classical shadows estimator).

In a fully Bayesian approach, one would instead seek the posterior expectation of $f(\rho)$ given the observed data. That is, the ideal Bayesian estimator is the conditional expectation of $f(\rho)$ under the posterior distribution of $\rho$ after observing all $N$ measurement outcomes. Computing this exact posterior expectation is generally intractable for large quantum systems due to two fundamental challenges. First, it requires performing continuous integrations over state spaces. Second, the dimension of the quantum state \(\rho\) grows exponentially with the number of qubits. In the next section, we describe our learning-based framework for approximating the Bayesian estimator, which enables the incorporation of arbitrary priors without explicitly evaluating a high-dimensional integral.

\section{Methods}
\label{sec:methods}

\subsection{PI network and residual learning}
\label{sec:pi_model_description}

Our approach is to train a neural network to estimate
\begin{equation}
\label{equation:posterior_sequence}
\E \left[ f(\rho) \mid \left( \mathbf{v}^{(1)}, \dots , \mathbf{v}^{(N)} \right) \right],
\end{equation}
the Bayesian posterior expectation of $f(\rho)$ using the measurement outcomes as input. Instead of directly predicting $f(\rho)$, we find it advantageous to leverage any existing estimator $g$ (such as the classical shadows estimator) and train the network to predict the residual error relative to this estimator. In other words, the network learns to predict the difference between the current estimator's output $\hat{f}(\rho)$ and the true value $f(\rho)$. By correcting this difference, the network effectively learns to output a refined estimate closer to the ideal Bayesian value. That is, we train the model to predict the expected estimation error conditioned on all $N$ measurement results.

Even if each measurement choice can depend on all previous outcomes, the final posterior depends only on the multiset of observed values. Formally, we have the following observation.
\begin{observation}
\label{observation:permutation_invariance}
For any (possibly adaptive) measurement protocol in which the choice of the \(k\)-th measurement may depend (possibly at random) only on the previous outcomes \(\left(\mathbf{v}^{(1)}, \dots , \mathbf{v}^{(k-1)}\right)\), (\ref{equation:posterior_sequence}) is invariant under permutations of the measurement data \( \left( \mathbf{v}^{(1)}, \dots , \mathbf{v}^{(N)} \right) \).
\end{observation}
\begin{proof}
See Appendix~\ref{sec:permutation_invariance_proof}.
\end{proof}
We can therefore express the conditional expectation in terms of the multiset of outcomes, $\left\{\mathbf{v}^{(k)}\right\}_{k=1}^N$:
\begin{equation}
\label{equation:posterior_set}
\E \left[ \hat{f}(\rho) - f(\rho) \mid \left\{ \mathbf{v}^{(k)} \right\}_{k=1}^N \right].
\end{equation}
Our model is specifically designed to capture this permutation invariance. We employ the set transformer architecture \citep{lee2019set}, a neural network tailored for set-input problems, to process the collection of measurement outcome features. The set transformer uses a self-attention mechanism to produce a PI representation of the input set. They have been shown to be universal approximators for PI functions on sets. Following its original implementation, the network processes the input
\[
V = \begin{bmatrix}
\mathbf{v}^{(1)T}\\
\vdots\\
\mathbf{v}^{(N)T}
\end{bmatrix} \in \R^{N\times d}
\]
through two \emph{induced set attention block} (ISAB) layers (each with a fixed number of \emph{inducing points}), followed by a \emph{pooling by multi-head attention} (PMA) layer that aggregates the information across the $N$ inputs into a fixed-size representation. This is followed by a couple of \emph{set attention blocks} (SAB) operating on the pooled representation, and finally a linear layer that outputs a single real number (see FIG.~\ref{fig:Set_Transformer_Architecture} in Appendix~\ref{sec:set_transformer_diagram}). The output of the network can be interpreted as a prediction for the quantity (\ref{equation:posterior_set}).

During training, the mean squared error (MSE) between the network's prediction and the true target value $f(\rho)$ (which is known for training examples) is minimized, since this is equivalent to computing the posterior mean of \(f(\rho)\). By learning the residual relative to the baseline estimator, the network focuses on the less challenging part of the estimation task (e.g., correcting any bias due to finite $N$ and prior information) while inheriting the baseline's performance for scalar functions.

We also describe a calibration mechanism for our learned estimator. Specifically, we define a function $\sigma(x, F)$ that combines the raw estimator $F = \hat{f}(\rho)$ with the network’s learned correction in a controlled way. The parameter $x \in \R$ governs the degree to which the correction is applied. Suppose \(f_\text{l} \le f(\rho) \le f_\text{u}\) holds regardless of \(\rho\). Then one can define
\[
\sigma (x,F) = F_\text{c} \times (1-\tanh |x|) + \mathbf{1}_{\mathbb{R}^+}(x)\tanh x,
\]
where $F_\text{c} = \max ( \min ( F, f_\text{u} ), f_\text{l} )$ is a clipped estimate and \(\mathbf{1}_A(x)=[x\in A]\) is the indicator function. At $x=0$, we have $\sigma(0,F) = F_\text{c}$.

Alternatively, one could learn the estimator directly in the form \(\text{Sigmoid}(x)\), rather than using a residual correction. However, as shown in the experimental results, this direct approach is outperformed by classical shadows as the number of measurements \(N\) increases. This is because inverting the shadow requires learning parity-like functions that conventional learning algorithms notoriously struggle to approximate \citep{bengio2005curse, eldan2016power, telgarsky2016benefits, daniely2020learning}.

\subsection{Noisy measurements}
A common benchmark in simulation experiments is the measurement bit-flip noise model, where each qubit measurement outcome is independently flipped with probability \(\lambda\).
\begin{observation}
\label{observation:bit_flip_noise}
Under bit-flip noise with parameter \(\lambda\neq\frac{1}{2}\), the Bayesian posterior over \(\rho\) converges in distribution to a point mass at the true state in the limit of infinitely many measurements.
\end{observation}
\begin{proof}
See Appendix~\ref{sec:bit_flip_noise_proof}.
\end{proof}
As implied by Observation~\ref{observation:bit_flip_noise}, while the shadow protocol may not converge under nonzero noise, our model is, in principle, capable of recovering the true value given sufficiently many measurement records. A similar argument applies to any invertible noise model.

\subsection{Encoding of measurement statistics}

Each single-round measurement outcome is encoded as a feature vector \(\mathbf{v}^{(k)} \in \Z^d\), where
\[
d = 
\begin{cases}
n & \text{for random Pauli measurements}\\
n(2n+3) & \text{for random Clifford measurements}.
\end{cases}
\]

\paragraph{Pauli measurements.}For random Pauli measurements, each qubit's measurement basis is drawn from \(\{X,Y,Z\}\) and encoded as an integer via
\[
X\mapsto 0,\quad Y\mapsto 2,\quad Z\mapsto 4.
\]
Let \(\hat{\mathbf{b}}\in\{0,1\}^n\) be the observed outcome bits, and let \(\mathbf{p}\in\{0,2,4\}^n\) be the basis‑encoding vector. We then form \(v^{(k)} = \mathbf{p} + \hat{\mathbf{b}}\). For example, measuring a 3-qubit Pauli string \(XZY\) and obtaining outcome \(\hat{\mathbf{b}}=(1,0,1)\) yields \(\mathbf{v}^{(k)} = (1,4,3)\). Stacking \(N\) such vectors produces an \(N\times n\) feature matrix (see FIG.~\ref{fig:Pauli_encoding} in Appendix~\ref{sec:encoding_diagram}).

\paragraph{Clifford measurements.}Each Clifford circuit is specified by its stabilizer tableau of size \(2n\times(2n+1)\) \citep{dehaene2003clifford, aaronson2004improved, bravyi2019simulation}.  Concretely, each of the \(2n\) rows consists of:
\begin{itemize}
  \item \(n\) bits encoding the presence of \(X\) on each qubit,
  \item \(n\) bits encoding the presence of \(Z\) on each qubit,
  \item one phase bit in \(\{0,1\}\) representing eigenvalues in \(\{-1,+1\}\).
\end{itemize}
We do not delve into the details of stabilizer formalism here; it suffices to note that the above binary encoding fully specifies the measurement setting. Pauli symbols are encoded as
\[
\I\mapsto0,\quad X\mapsto1,\quad Y\mapsto2,\quad Z\mapsto3,
\]
and phases are encoded as
\[
-1\mapsto0,\quad +1\mapsto1.
\]
Flattening the tableau encoding and concatenating \(\hat{\mathbf{b}}\) yields a vector in \(\mathbb{Z}^{n(2n+3)}\).  Stacking \(N\) such vectors produces an \(N\times n(2n+3)\) feature matrix (see FIG.~\ref{fig:Clifford_encoding} in Appendix~\ref{sec:encoding_diagram}).

\paragraph{Variable-length measurement data.}Thus far, it was assumed that the number of measurements \(N\) was fixed. To enable the use of a pretrained model across varying numbers of measurements, a padding-based strategy is adopted to ensure uniform input dimensionality. Specifically, we fix the maximum number of measurement rounds to \(N_{\max} = 100\) during training. For a measurement sequence of length \(N < N_{\max}\), an input tensor of shape \(N \times d\) is constructed and a padding tensor of shape \((N_{\max} - N) \times d\) with all entries set to \(-1\) is appended (note that in this setup, the padding value \(-1\) is guaranteed to lie outside the vocabulary of valid measurement statistics). This results in a fixed-size input of shape \(N_{\max} \times d\). During training, the model learns to recognize and ignore these padded entries, treating them as invalid or non-informative statistics.

\subsection{Application to fidelity estimation}
\label{sec:application_dfe}

As a first demonstration, we consider direct fidelity estimation (DFE), the task of estimating the fidelity of the unknown state $\rho$ with a target entangled state. Specifically, we take the target state to be the $n$-qubit Greenberger-Horne-Zeilinger (GHZ) state \citep{greenberger1989going, greenberger1990bell} defined as
\[
\ket{\psi_\text{GHZ}} = \frac{1}{\sqrt{2}} \left( 
\ket{0}^{\otimes n} + \ket{1}^{\otimes n} \right).
\]
The fidelity with the GHZ state is $f(\rho) = \langle \psi_{\mathrm{GHZ}}|\rho|\psi_{\mathrm{GHZ}}\rangle$. There exists an efficient procedure for estimating the fidelity of an unknown quantum state with the GHZ state \citep{cha2025efficient}. Algorithm~\ref{alg:ghz_dfe} in Appendix~\ref{sec:deferred_algorithms} describes this procedure. This baseline procedure is referred to as the ``Shadow'' estimator for fidelity. The estimator in Algorithm~\ref{alg:ghz_dfe} has the following desirable properties.

\begin{observation}
\label{observation:dfe_estimator_variance}
In Algorithm~\ref{alg:ghz_dfe},
\begin{equation}
\label{equation:var_F}
\textnormal{Var}(F)=\frac{(1+2f(\rho))(1-f(\rho))}{2}.
\end{equation}
In particular,
\[
\lim_{\rho \rightarrow \ketbra{\psi_\textnormal{GHZ}}}\textnormal{Var}(F) = \lim_{f(\rho) \rightarrow 1}\textnormal{Var}(F) = 0.
\]
\end{observation}

\begin{proof}
See Appendix~\ref{sec:dfe_estimator_variance_proof}.
\end{proof}

\begin{lemma}[Informal]
\label{lemma:algorithm_locally_optimal}
Algorithm~\ref{alg:ghz_dfe} is locally optimal in the sense that, within the class of adaptive strategies that sample from the same set of measurement settings with reweighted estimators, no further improvement is possible.

\end{lemma}
\begin{proof}
See Appendix~\ref{sec:algorithm_locally_optimal_proof}.
\end{proof}

From Observation~\ref{observation:dfe_estimator_variance}, it can be seen that if the fidelity is close to 1, then Algorithm~\ref{alg:ghz_dfe} itself (non-Bayesian) is quite accurate. Nevertheless, as demonstrated in our experimental results, the proposed model consistently improves this estimator through Bayesian inference.

\subsection{Application to entanglement entropy estimation}
\label{sec:application_entropy}

As a second example, we consider the task of estimating a non-linear function of the state: the second-order R\'enyi entanglement entropy across a bipartition of the qubits \citep{hastings2010measuring, islam2015measuring, brydges2019probing}. Let the qubits be partitioned into subset $A$ and its complement $B$ (for instance, $A$ could be the first $n/2$ qubits and $B$ the remaining $n/2$ qubits). The quantity of interest can be expressed as
\[
f(\rho) = \tr((\rho \otimes \rho) S_A),
\]
where $S_A$ is the local swap operator acting on two copies of subsystem $A$ (the R\'enyi entropy is then $-\log f(\rho)$, but here we focus on estimating $f(\rho)$ itself). In other words, the swap operator $S_A$ acts on the two-copy Hilbert space by exchanging the states of subsystem $A$ between the two copies, while acting trivially on subsystem $B$. In the computational basis, its action can be written as
\[
S_A (\ket{i}_A \ket{j}_B \ket{k}_A \ket{l}_B) = \ket{k}_A \ket{j}_B \ket{i}_A \ket{l}_B,
\]
for any basis states $\ket{i}_A,\ket{k}_A$ of subsystem $A$ and $\ket{j}_B,\ket{l}_B$ of subsystem $B$. While $f(\rho)$ is a non-linear function of $\rho$, it can be efficiently estimated using the classical shadows framework \citep{huang2020predicting}. Algorithm~\ref{alg:renyi} in Appendix~\ref{sec:deferred_algorithms} describes this procedure. This baseline procedure is referred to as the ``Shadow'' estimator for entanglement entropy.

\begin{observation}
\label{observarion:renyi_pauli_snapshot_is_local}
For random Pauli measurements, line~\ref{line:renyi_snapshot} in Algorithm~\ref{alg:renyi} reduces to the average of local snapshots of \(\tr(\rho_A^2)\).
\end{observation}

\begin{proof}
See Appendix~\ref{sec:renyi_pauli_snapshot_is_local_proof}.
\end{proof}

\section{Experimental results}
\label{sec:experiment_results}

\subsection{Model setup}
\label{sec:model_setup}

The proposed approach was evaluated on the two tasks described in sections~\ref{sec:application_dfe} \& \ref{sec:application_entropy} and compared to the classical shadows baseline. Our network was implemented as described in Section~\ref{sec:pi_model_description} with a hidden dimension of $d_h=128$, $4$ attention heads, and $32$ inducing points for each ISAB layer. The input feature dimension was determined by the encoding scheme for each shadow protocol. For each task, a training dataset of 10,000 instances and a test dataset of 1,000 instances were generated, where each instance consists of \(N\) repeated measurements. The model was trained using the Adam optimizer with a learning rate of $10^{-4}$ and a batch size of 50, for 10 epochs. Experiments were run on a machine with a 32-core Intel Xeon Processor (Skylake, IBRS) @ 2.5 GHz, equipped with an NVIDIA H100 NVL GPU, CUDA 12.9. As this is the first Bayesian tomography scheme that avoids exponential runtime, the only directly comparable baseline is the classical shadows protocol.

\begin{figure}[h]
    \centering

    \includegraphics[width=0.15\linewidth]{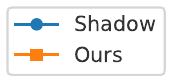}
    \par\vspace{1em}  

    \begin{subfigure}[b]{0.49\linewidth}
        \centering
        \includegraphics[page=1, width=\linewidth]{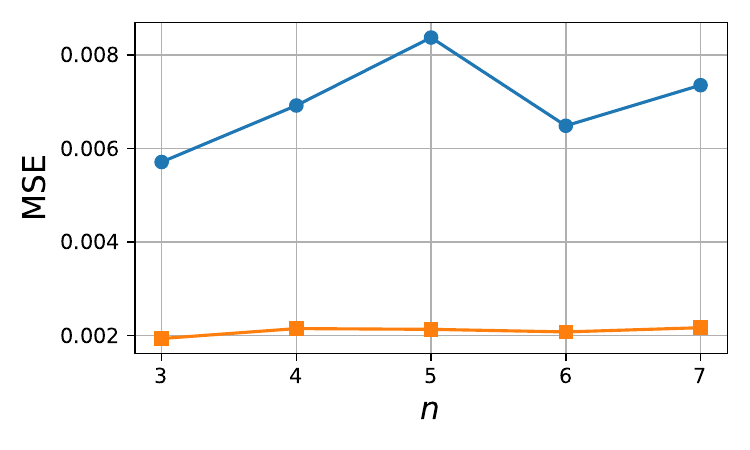}
        \caption{}
        \label{fig:DFE_GHZ_Pauli_N10}
    \end{subfigure}
    \hfill
    \begin{subfigure}[b]{0.49\linewidth}
        \centering
        \includegraphics[page=1, width=\linewidth]{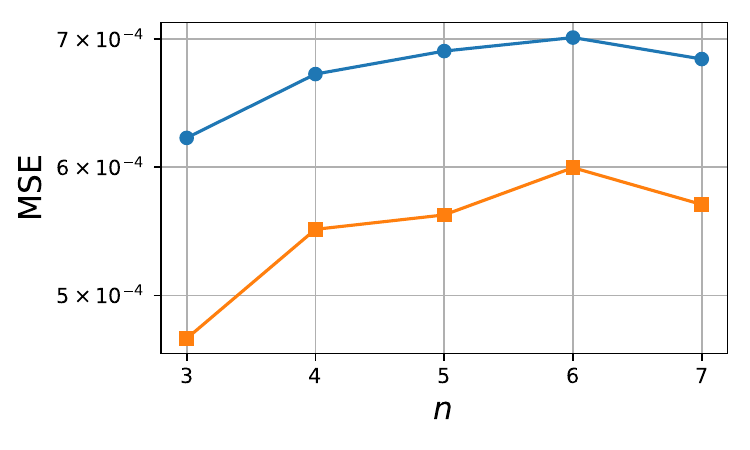}
        \caption{}
        \label{fig:DFE_GHZ_Pauli_N100}
    \end{subfigure}
    \begin{subfigure}[b]{0.49\linewidth}
        \centering
        \includegraphics[page=1, width=\linewidth]{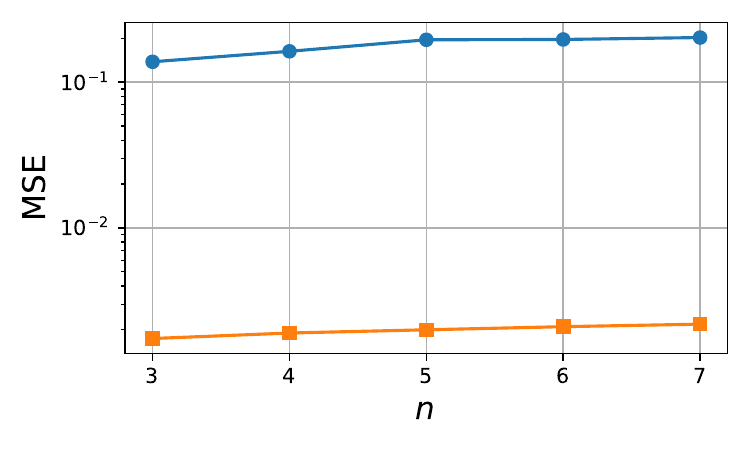}
        \caption{}
        \label{fig:DFE_GHZ_Clifford_N10}
    \end{subfigure}
    \hfill
    \begin{subfigure}[b]{0.49\linewidth}
        \centering
        \includegraphics[page=1, width=\linewidth]{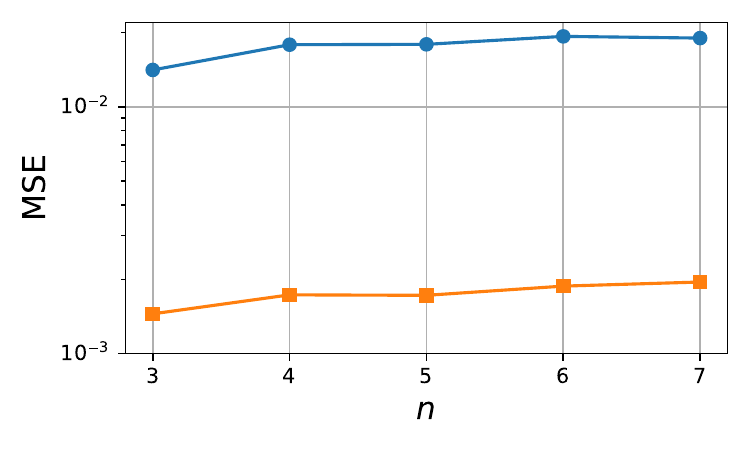}
        \caption{}
        \label{fig:DFE_GHZ_Clifford_N100}
    \end{subfigure}
    \caption{\justifying MSE comparison between classical shadow estimation and the proposed Bayesian estimation. Displayed are DFE results for the GHZ state using (a) random Pauli measurements with \(N=10\), (b) random Pauli measurements with \(N=100\), (c) random Clifford measurements with \(N=10\), and (d) random Clifford measurements with \(N=100\).}
    \label{fig:DFE_GHZ_Pauli_Clifford_comparison}
\end{figure}

\begin{figure}[h]
    \centering

    \includegraphics[width=0.15\linewidth]{plots/legend_box.pdf}
    \par\vspace{1em}  

    \begin{subfigure}[b]{0.49\linewidth}
        \centering
        \includegraphics[page=1, width=\linewidth]{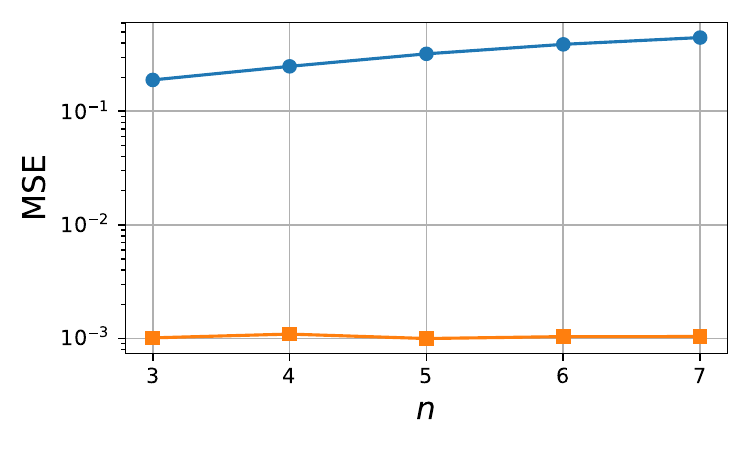}
        \caption{}
        \label{fig:DFE_GHZ_Pauli_N10_BitFlip}
    \end{subfigure}
    \hfill
    \begin{subfigure}[b]{0.49\linewidth}
        \centering
        \includegraphics[page=1, width=\linewidth]{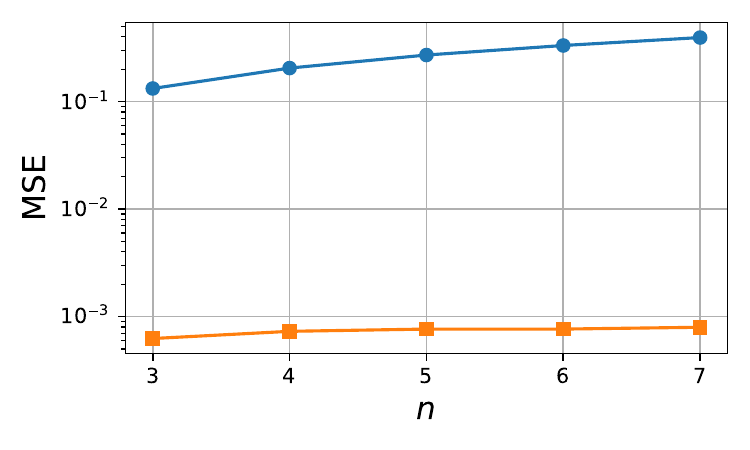}
        \caption{}
        \label{fig:DFE_GHZ_Pauli_N100_BitFlip}
    \end{subfigure}
    \begin{subfigure}[b]{0.49\linewidth}
        \centering
        \includegraphics[page=1, width=\linewidth]{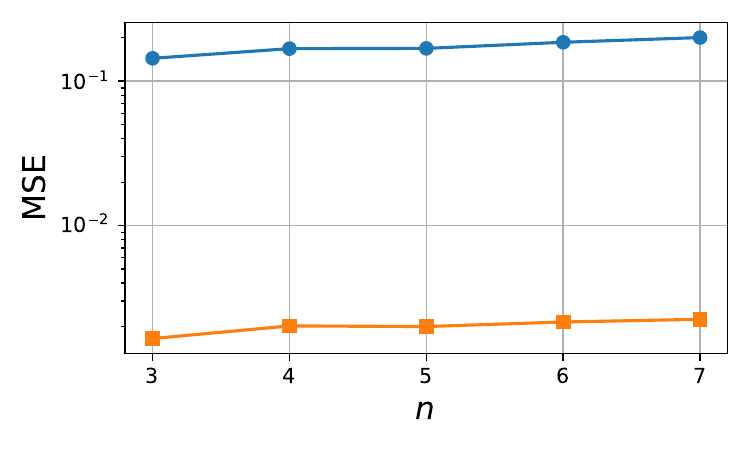}
        \caption{}
        \label{fig:DFE_GHZ_Clifford_N10_BitFlip}
    \end{subfigure}
    \hfill
    \begin{subfigure}[b]{0.49\linewidth}
        \centering
        \includegraphics[page=1, width=\linewidth]{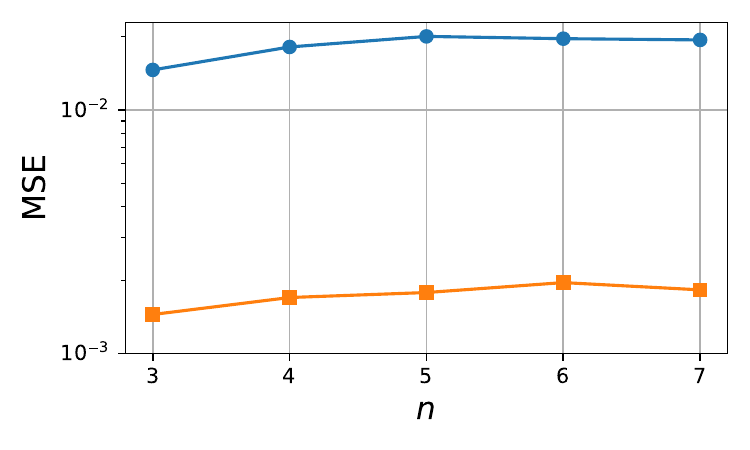}
        \caption{}
        \label{fig:DFE_GHZ_Clifford_N100_BitFlip}
    \end{subfigure}
    \caption{\justifying MSE comparison between classical shadow estimation and the proposed Bayesian estimation under bit-flip noise with parameter \(\lambda=0.1\). Displayed are DFE results for the GHZ state using (a) random Pauli measurements with \(N=10\), (b) random Pauli measurements with \(N=100\), (c) random Clifford measurements with \(N=10\), and (d) random Clifford measurements with \(N=100\).}
    \label{fig:DFE_GHZ_Pauli_Clifford_BitFlip_comparison}
\end{figure}

\begin{figure}[h]
    \centering

    \includegraphics[width=0.15\linewidth]{plots/legend_box.pdf}
    \par\vspace{1em}  
    
    \begin{subfigure}[b]{0.49\linewidth}
        \centering
        \includegraphics[page=1, width=\linewidth]{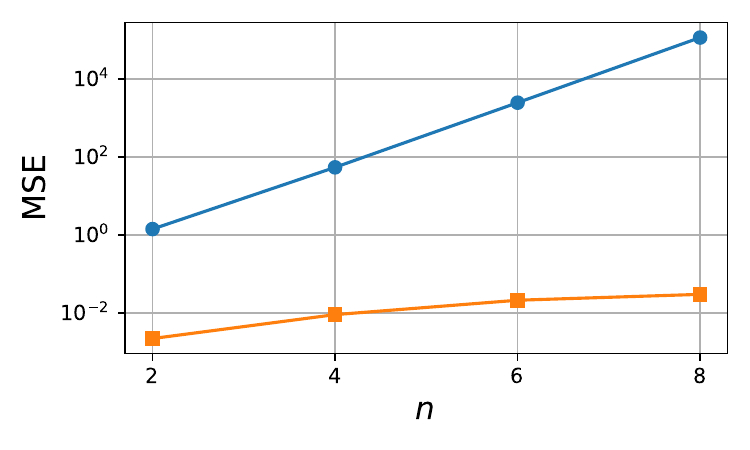}
        \caption{}
        \label{fig:Renyi_Pauli_N10}
    \end{subfigure}
    \hfill
    \begin{subfigure}[b]{0.49\linewidth}
        \centering
        \includegraphics[page=1, width=\linewidth]{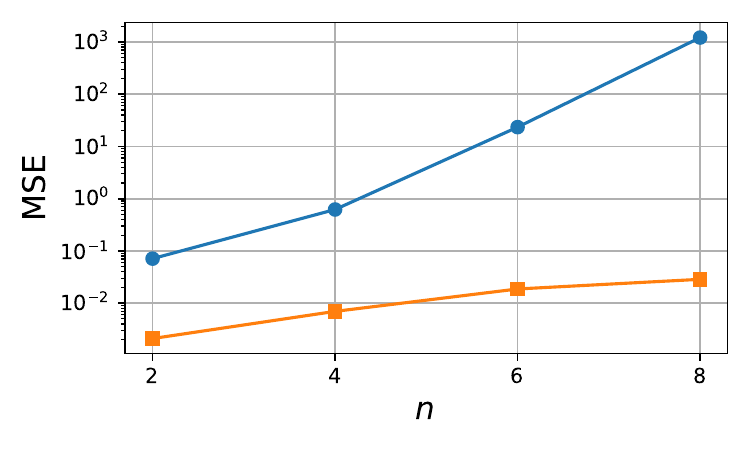}
        \caption{}
        \label{fig:Renyi_Pauli_N100}
    \end{subfigure}
    \begin{subfigure}[b]{0.49\linewidth}
        \centering
        \includegraphics[page=1, width=\linewidth]{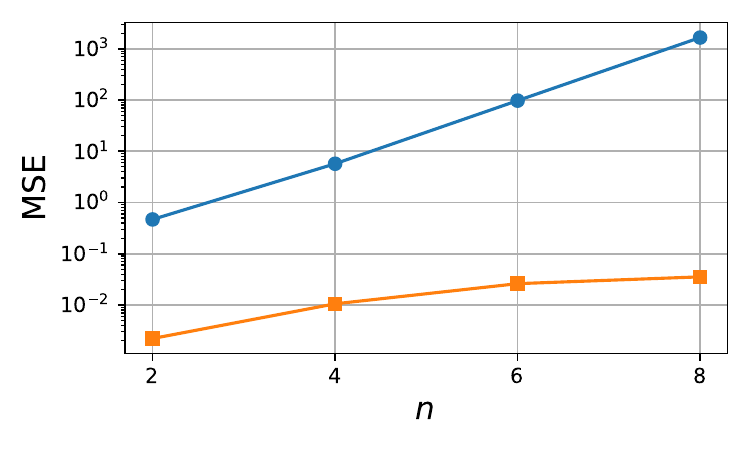}
        \caption{}
        \label{fig:Renyi_Clifford_N10}
    \end{subfigure}
    \hfill
    \begin{subfigure}[b]{0.49\linewidth}
        \centering
        \includegraphics[page=1, width=\linewidth]{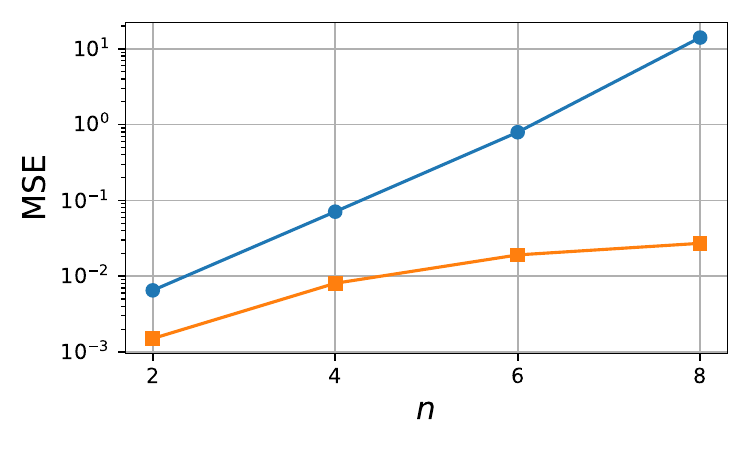}
        \caption{}
        \label{fig:Renyi_Clifford_N100}
    \end{subfigure}
    \caption{\justifying MSE comparison between classical shadow estimation and the proposed Bayesian estimation. Displayed are entanglement entropy estimation results for the GHZ state using (a) random Pauli measurements with \(N=10\), (b) random Pauli measurements with \(N=100\), (c) random Clifford measurements with \(N=10\), and (d) random Clifford measurements with \(N=100\).}
    \label{fig:Renyi_Pauli_Clifford_comparison}
\end{figure}

\begin{figure}[h]
    \centering

    \includegraphics[width=0.15\linewidth]{plots/legend_box.pdf}
    \par\vspace{1em}  

    \begin{subfigure}[b]{0.49\linewidth}
        \centering
        \includegraphics[page=1, width=\linewidth]{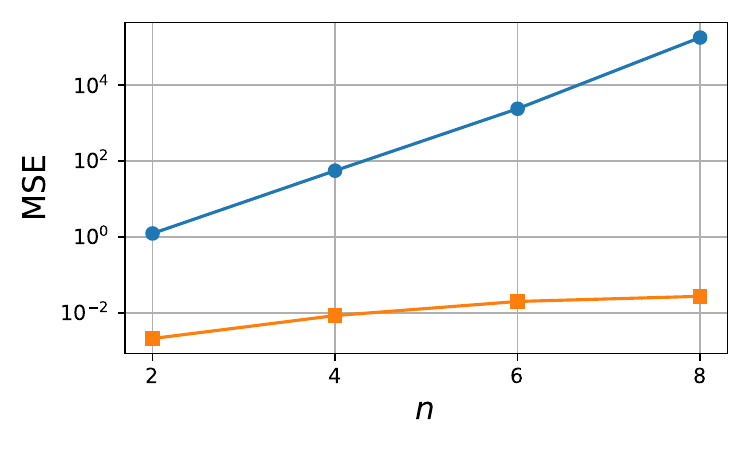}
        \caption{}
        \label{fig:Renyi_Pauli_N10_BitFlip}
    \end{subfigure}
    \hfill
    \begin{subfigure}[b]{0.49\linewidth}
        \centering
        \includegraphics[page=1, width=\linewidth]{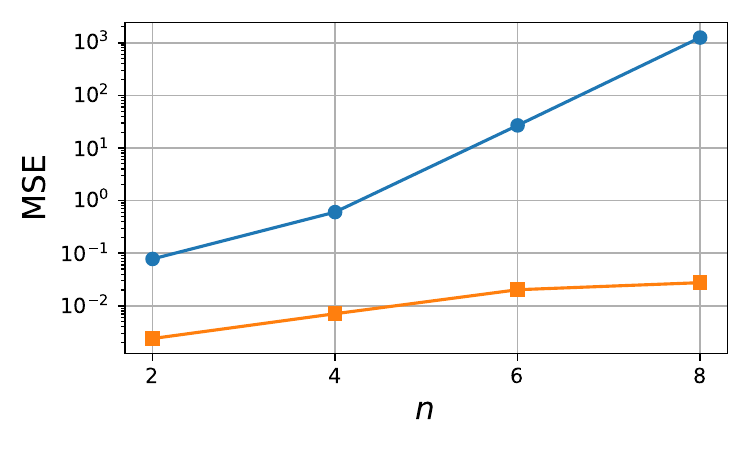}
        \caption{}
        \label{fig:Renyi_Pauli_N100_BitFlip}
    \end{subfigure}
    \begin{subfigure}[b]{0.49\linewidth}
        \centering
        \includegraphics[page=1, width=\linewidth]{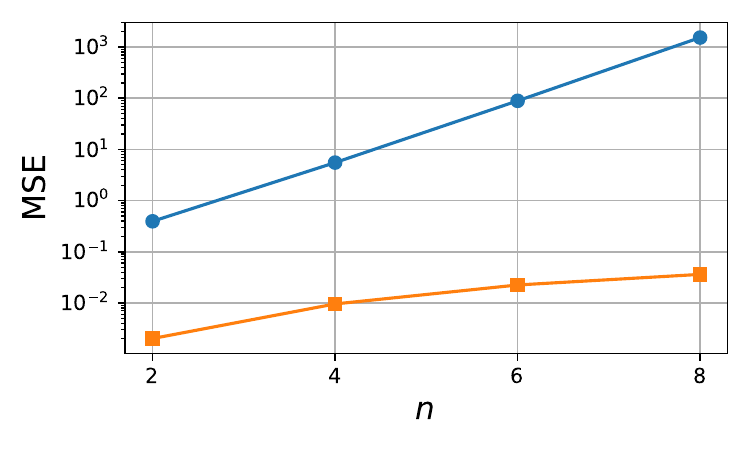}
        \caption{}
        \label{fig:Renyi_Clifford_N10_BitFlip}
    \end{subfigure}
    \hfill
    \begin{subfigure}[b]{0.49\linewidth}
        \centering
        \includegraphics[page=1, width=\linewidth]{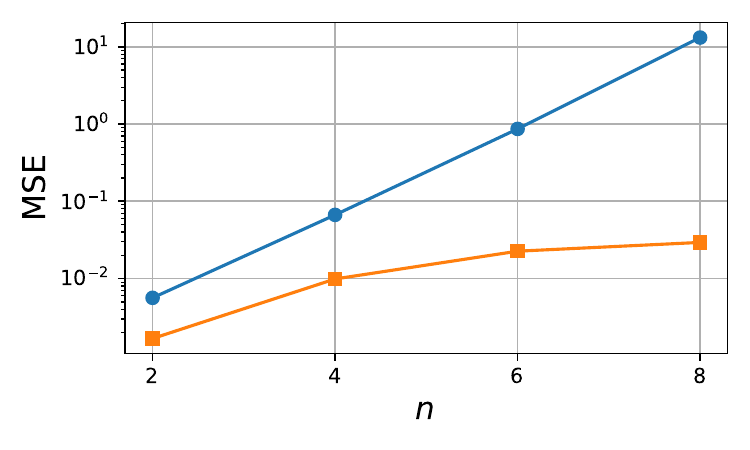}
        \caption{}
        \label{fig:Renyi_Clifford_N100_BitFlip}
    \end{subfigure}
    \caption{\justifying MSE comparison between classical shadow estimation and the proposed Bayesian estimation under bit-flip noise with parameter \(\lambda=0.1\). Displayed are entanglement entropy estimation results for the GHZ state using (a) random Pauli measurements with \(N=10\), (b) random Pauli measurements with \(N=100\), (c) random Clifford measurements with \(N=10\), and (d) random Clifford measurements with \(N=100\).}
    \label{fig:Renyi_Pauli_Clifford_BitFlip_comparison}
\end{figure}

\begin{figure}[h]
    \centering
    \begin{subfigure}[b]{0.49\linewidth}
        \centering
        \includegraphics[page=1, width=\linewidth]{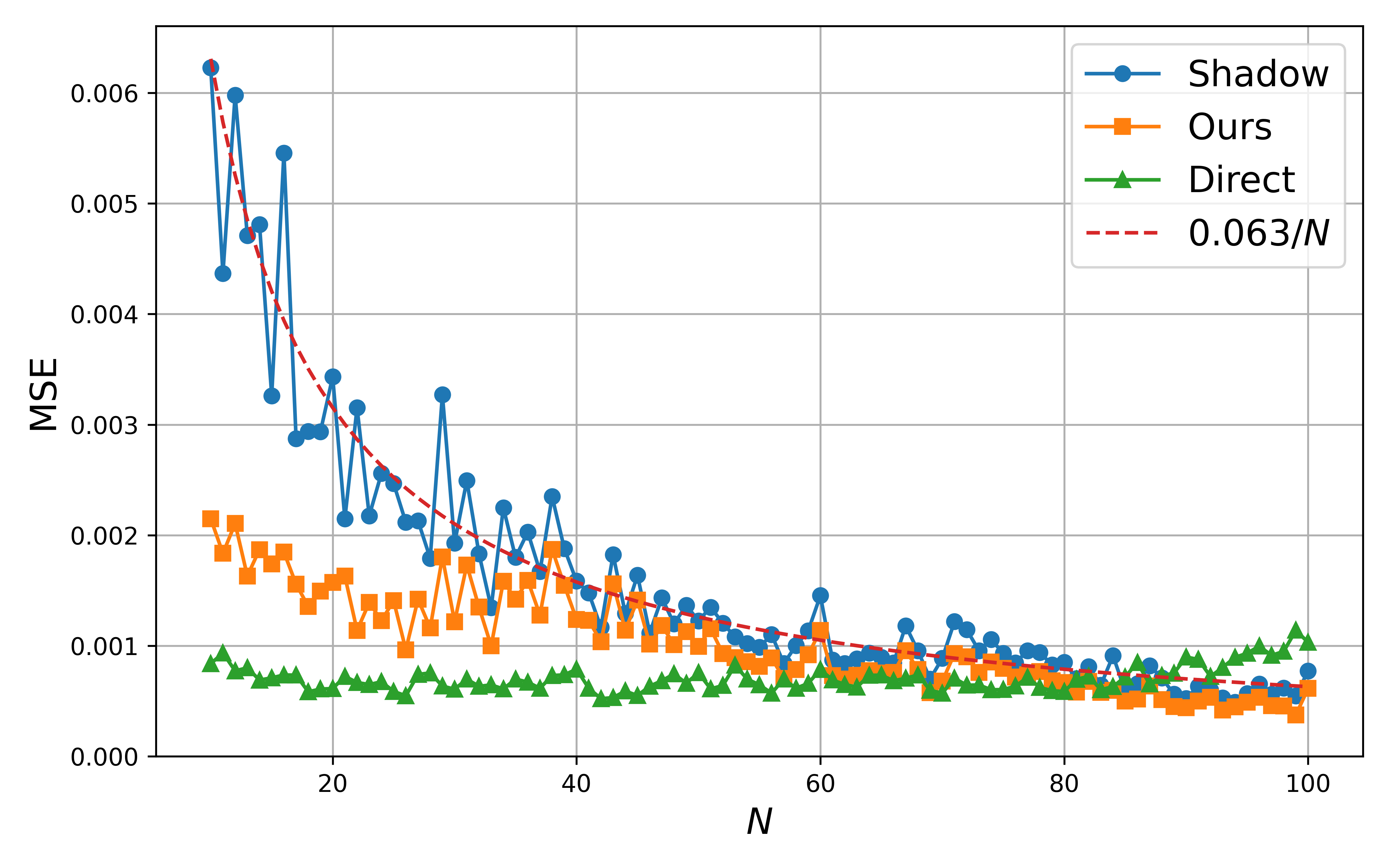}
        \caption{}
        \label{fig:varying_N_avg_DFE_Pauli}
    \end{subfigure}
    \hfill
    \begin{subfigure}[b]{0.49\linewidth}
        \centering
        \includegraphics[page=1, width=\linewidth]{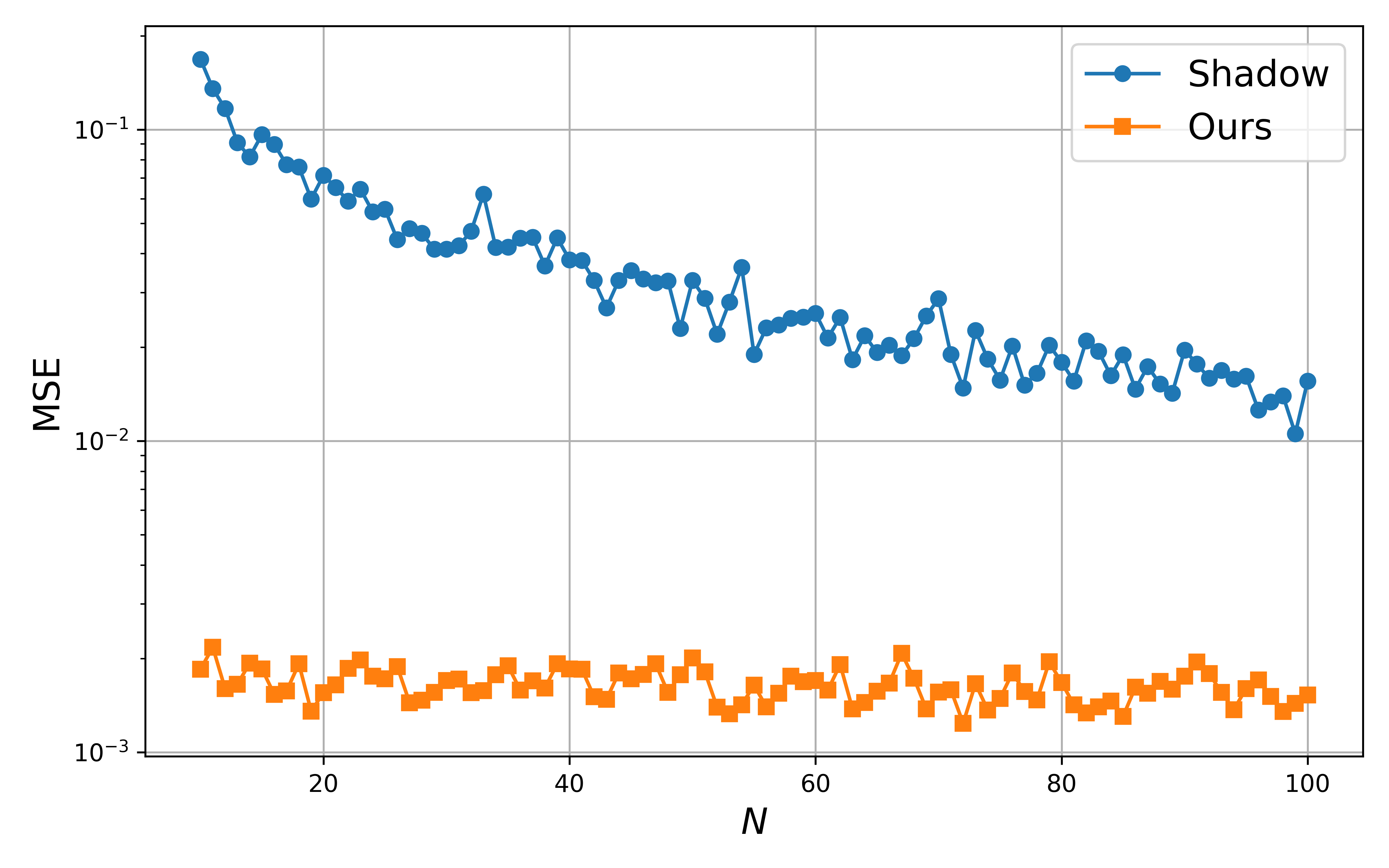}
        \caption{}
        \label{fig:varying_N_avg_DFE_Clifford}
    \end{subfigure}
    \caption{\justifying MSE comparison between classical shadow estimation and the proposed Bayesian estimation. Displayed are DFE results for the 3-qubit GHZ state with varying number of measurements \(N\) using (a) random Pauli measurements and (b) random Clifford measurements. In (a), \emph{Direct} refers to the non-residual learning, and the dashed line represents the theoretical expected variance of the classical shadow estimator.}
    \label{fig:varying_N_DFE}
\end{figure}

\begin{figure}[h]
    \centering
    \begin{subfigure}[b]{0.49\linewidth}
        \centering
        \includegraphics[page=1, width=\linewidth]{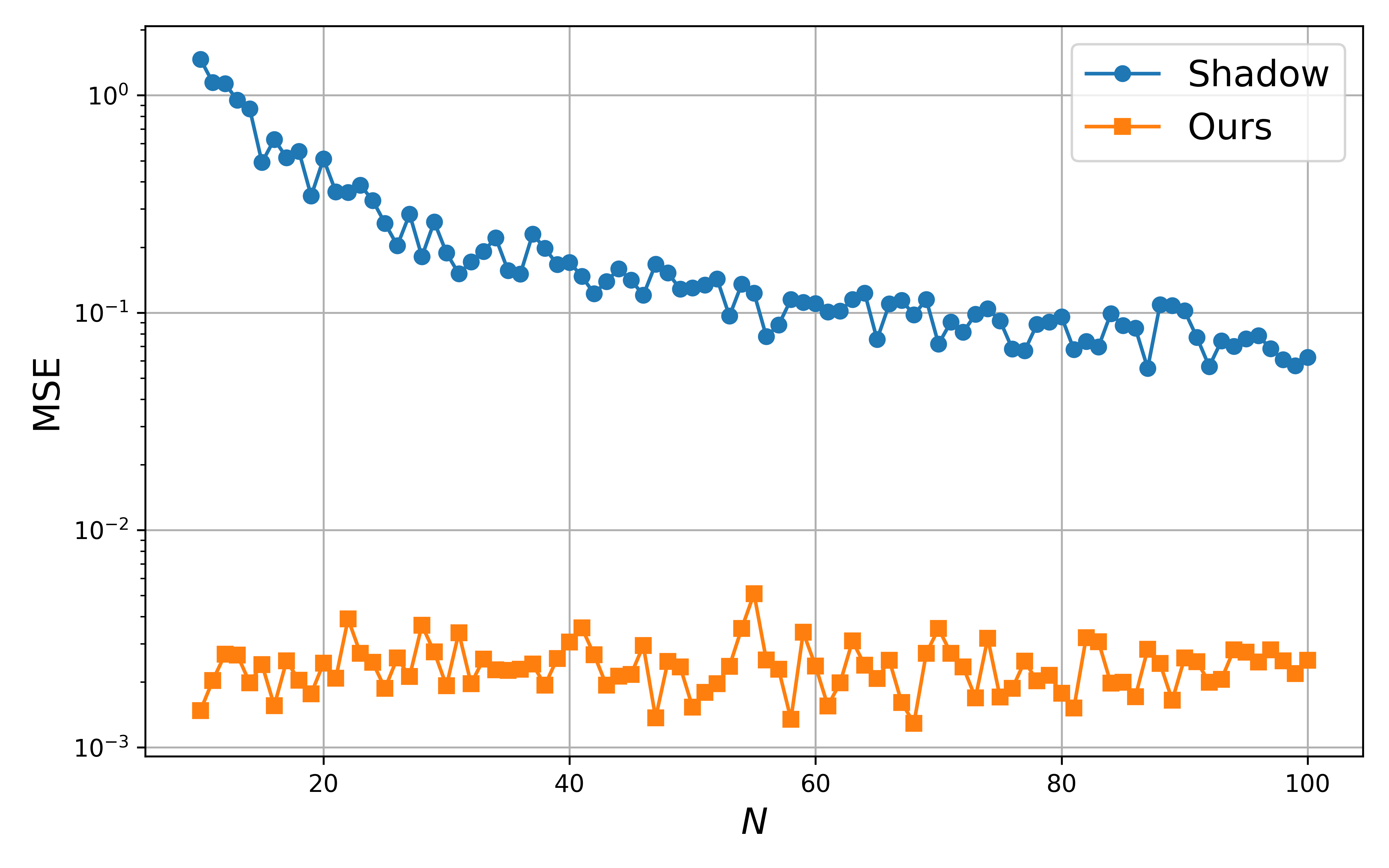}
        \caption{}
        \label{fig:varying_N_avg_Renyi_Pauli}
    \end{subfigure}
    \hfill
    \begin{subfigure}[b]{0.49\linewidth}
        \centering
        \includegraphics[page=1, width=\linewidth]{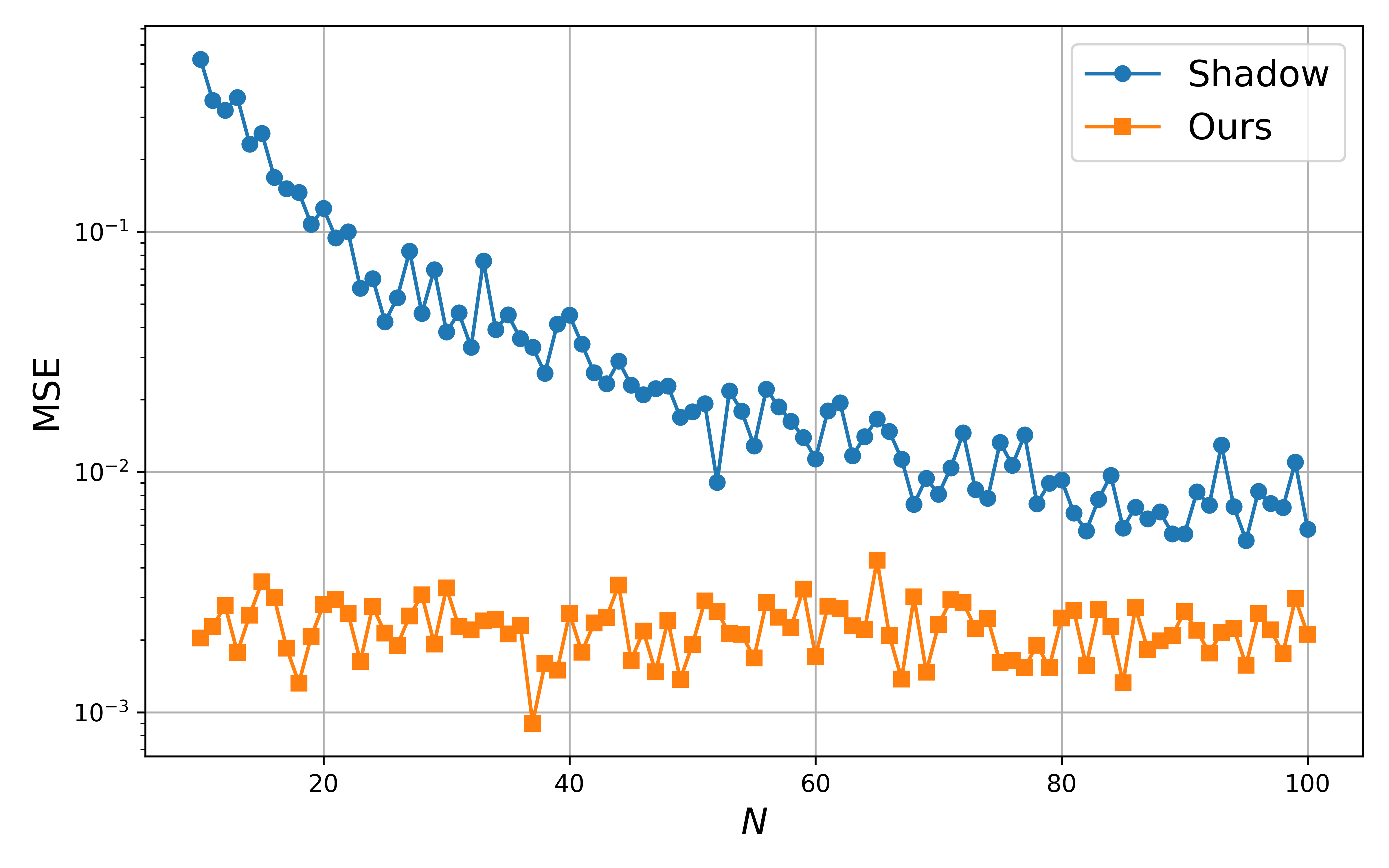}
        \caption{}
        \label{fig:varying_N_avg_Renyi_Clifford}
    \end{subfigure}
    \caption{\justifying MSE comparison between classical shadow estimation and the proposed Bayesian estimation. Displayed are entanglement entropy estimation results for the 2-qubit GHZ state (i.e., the Bell state \(|\Phi^+\rangle\)) with varying number of measurements \(N\) using (a) random Pauli measurements and (b) random Clifford measurements.}
    \label{fig:varying_N_Renyi}
\end{figure}

\subsection{DFE for the GHZ state}

For GHZ fidelity, each instance consisted of a random $n$-qubit state along with $N$ simulated measurement outcomes according to Algorithm~\ref{alg:ghz_dfe}, and the corresponding ground truth fidelity $f(\rho)=\langle\psi_{\mathrm{GHZ}}|\rho|\psi_{\mathrm{GHZ}}\rangle$. The input state was sampled by applying a depolarizing channel to the GHZ state:
\[
\rho = (1-\lambda)\ketbra{\psi_\text{GHZ}} + \frac{\lambda}{2^n}\I^{\otimes n},
\]
where the depolarizing noise parameter \(\lambda\) was drawn uniformly from the interval \((0,0.1)\). If \(n=3\), the fidelity
\[
\langle\psi_{\mathrm{GHZ}}|\rho|\psi_{\mathrm{GHZ}}\rangle = 1-\lambda\left(1-\frac{1}{2^n}\right)
\]
is uniformly distributed over the interval \((0.9125,1)\) and in Algorithm~\ref{alg:ghz_dfe},
\begin{equation}
\E_\rho[\text{Var}(\hat{F})] = \frac{\E_\rho[\text{Var}(F)]}{N} \approx \frac{0.063}{N}.
\end{equation}
This curve is plotted in FIG.~\ref{fig:varying_N_avg_DFE_Pauli} for reference.

\subsection{Second-order R\'enyi entanglement entropy estimation for the GHZ state}

For R\'enyi entropy, each instance consists of a random $n$-qubit state (with a chosen bipartition $A|B$) along with $N$ simulated measurement outcomes according to Algorithm~\ref{alg:renyi}, and the corresponding ground truth entanglement entropy $f(\rho)=\tr((\rho\otimes\rho)S_A)$. The input state was the GHZ state, but an adversary randomly replaces it with a spurious state \(\omega\), sampled according to the Hilbert-Schmidt metric \citep{zyczkowski2001induced}. The replacement occurred with probability \(p\) drawn uniformly from the interval \((0,1)\):
\[
\rho = (1-p)\ketbra{\psi_\text{GHZ}} + p\omega.
\]

\subsection{Performance evaluation}

The model was trained to minimize MSE loss between its output \(\sigma(x,F)\) and the true $f(\rho)$ for each instance, using the Adam optimizer. We repeated training for system sizes $n=3,4,5,6,7$ for GHZ fidelity, and $n=2,4,6,8$ qubits for R\'enyi entropy (where subsystems $A$ and $B$ are equal-sized, each comprising \(n/2\) qubits). FIGS.~\ref{fig:DFE_GHZ_Pauli_Clifford_comparison}--\ref{fig:Renyi_Pauli_Clifford_BitFlip_comparison} summarize the results on the test sets. In all cases, the proposed Bayesian estimator achieves a lower MSE than the naive classical shadows estimator. Additionally, the model was trained on 10,000 experiments, with the number of measurements \(N\) uniformly sampled from the range 10 to 100 for each instance. For evaluation, 100 test experiments were generated for each fixed value of \(N\), and the corresponding MSEs are plotted in FIGS.~\ref{fig:varying_N_DFE} and \ref{fig:varying_N_Renyi}. The results show that our Bayesian estimator consistently outperforms the naive classical shadows estimator across the full range of evaluated measurement numbers.

\begin{remark}
FIG.~\ref{fig:varying_N_avg_DFE_Pauli} shows that direct learning of the target functional outperforms residual learning for small \(N\). In this regime, the variance of the classical shadow estimator is so large that a crude estimate drawn from the feasible range of \(f(\rho)\) may produce a better result. However, as \(N\) increases and the shadow estimator becomes sufficiently accurate, residual learning outperforms direct learning. This indicates that approximating the shadow inversion from measurement data to a point estimate of \(f(\rho)\) is much more challenging than learning only the bias of an existing estimator, since many shadow inversion steps include evaluations of parity-like functions (see Algorithm~\ref{alg:ghz_dfe} and Eq.~(\ref{equation:local_renyi_snapshot})).
\end{remark}

\begin{table}[ht]
  \centering
  \caption{\justifying Per-experiment inference time for the proposed Bayesian estimation, considering only the residual term and excluding the shadow estimation procedure.}
  \label{tab:times}
  \renewcommand{\arraystretch}{1.1} 
  \begin{minipage}[t]{0.48\textwidth}
    \centering
    {
    \setlength{\tabcolsep}{5pt}
    \begin{tabular}{| c | c | c | c | c |}
      \toprule
      \textbf{Task} & \textbf{Protocol} & $n$ & $N$ & \textbf{Time (\textmu s)} \\
      \midrule
      \multirow[c]{20}{*}{\textbf{DFE}} &
        \multirow[c]{10}{*}{\textbf{Pauli}} &
        \multirow[c]{2}{*}{3} & 10  & 2.204 \\
      & & & 100 & 4.314 \\
      & & \multirow[c]{2}{*}{4} & 10  & 2.236 \\
      & & & 100 & 4.303 \\
      & & \multirow[c]{2}{*}{5} & 10  & 2.275 \\
      & & & 100 & 4.357 \\
      & & \multirow[c]{2}{*}{6} & 10  & 2.277 \\
      & & & 100 & 4.370 \\
      & & \multirow[c]{2}{*}{7} & 10  & 2.250 \\
      & & & 100 & 4.306 \\
      \cmidrule(l){2-5}
      & \multirow[c]{10}{*}{\textbf{Clifford}} &
        \multirow[c]{2}{*}{3} & 10  & 2.201 \\
      & & & 100 & 4.395 \\
      & & \multirow[c]{2}{*}{4} & 10  & 2.258 \\
      & & & 100 & 4.432 \\
      & & \multirow[c]{2}{*}{5} & 10  & 2.396 \\
      & & & 100 & 4.491 \\
      & & \multirow[c]{2}{*}{6} & 10  & 2.225 \\
      & & & 100 & 4.487 \\
      & & \multirow[c]{2}{*}{7} & 10  & 2.185 \\
      & & & 100 & 4.515 \\
      \bottomrule
    \end{tabular}
    }
  \end{minipage}
  \hfill
  \begin{minipage}[t]{0.48\textwidth}
    \centering
    {
    \setlength{\tabcolsep}{5pt}
    \begin{tabular}{| c | c | c | c | c |}
      \toprule
      \textbf{Task} & \textbf{Protocol} & $n$ & $N$ & \textbf{Time (\textmu s)} \\
      \midrule
      \multirow[c]{16}{*}{\textbf{Entropy}} &
        \multirow[c]{8}{*}{\textbf{Pauli}} &
        \multirow[c]{2}{*}{2} & 10  & 2.236 \\
      & & & 100 & 4.305 \\
      & & \multirow[c]{2}{*}{4} & 10  & 2.244 \\
      & & & 100 & 4.295 \\
      & & \multirow[c]{2}{*}{6} & 10  & 2.206 \\
      & & & 100 & 4.287 \\
      & & \multirow[c]{2}{*}{8} & 10  & 2.239 \\
      & & & 100 & 4.317 \\
      \cmidrule(l){2-5}
      & \multirow[c]{8}{*}{\textbf{Clifford}} &
        \multirow[c]{2}{*}{2} & 10  & 2.260 \\
      & & & 100 & 4.331 \\
      & & \multirow[c]{2}{*}{4} & 10  & 2.274 \\
      & & & 100 & 4.409 \\
      & & \multirow[c]{2}{*}{6} & 10  & 2.348 \\
      & & & 100 & 4.464 \\
      & & \multirow[c]{2}{*}{8} & 10  & 2.279 \\
      & & & 100 & 4.549 \\
      \bottomrule
    \end{tabular}
    }
  \end{minipage}
\end{table}

Table~\ref{tab:times} provides the inference times per experiment. The runtimes reflect only the learned bias and do not include the classical shadow estimation itself, since optimizing the latter falls outside the scope of this work. As Table~\ref{tab:times} shows, the correction adds little overhead, so it can be seamlessly merged with existing estimation protocols.

\section{Discussion}
\label{sec:discussion}

\paragraph*{Limitations.}Our approach inherits a key limitation of data-driven methods; since the network is trained to learn the posterior correction of the classical shadow estimator, it requires a large training set in order to generalize reliably. Each training sample must be labeled with an almost exact value of the target property---i.e. the ground-truth posterior mean---which in practice can only be obtained by performing a very large number of measurements. Thus, although our architecture scales efficiently with the number of qubits and measurements, the cost of generating labels poses an obstacle when the variance of the shadow estimator grows with system size.

The problem of estimating properties of a quantum state in the presence of imperfect control and readout \citep{chen2021robust, koh2022classical, wu2024error} can be understood naturally within a Bayesian inference framework; one treats the parameters characterizing noisy unitaries and measurement channels as learnable latent variables, and then incorporates that information into posterior updates over the true quantum state.

Looking forward, one promising direction is to bypass the intermediate classical shadow reconstruction and instead update a posterior distribution directly over the target functional \(f(\rho)\). By framing the problem as inference on the scalar quantity \(f(\rho)\), one could potentially further reduce sample complexity and computational overhead. Another avenue for improvement lies in adaptive measurement strategies, where each subsequent measurement choice depends on previous outcomes to maximize information gain about \(\rho\) or \(f(\rho)\) \citep{quek2021adaptive, lange2023adaptive}. Adaptivity can offer performance gains, and permutationally invariant architectures can still be employed to process measurement records.

\section*{Acknowledgments}

This work is in part supported by the National Research Foundation of Korea (NRF, RS-2024-00451435 (20\%), RS-2024-00413957 (20\%)), Institute of Information \& communications Technology Planning \& Evaluation (IITP, RS-2021-II212068 (10\%), RS-2025-02305453 (15\%), RS-2025-02273157 (15\%), RS-2025-25442149 (10\%) RS-2021-II211343 (10\%)) grant funded by the Ministry of Science and ICT (MSIT), Institute of New Media and Communications (INMAC), and the BK21 FOUR program of the Education, Artificial Intelligence Graduate School Program (Seoul National University), and Research Program for Future ICT Pioneers, Seoul National University in 2025.

\bibliography{apssampv1}

@inproceedings{tang2024towards,
  title={Towards LLM4QPE: Unsupervised pretraining of quantum property estimation and a benchmark},
  author={Tang, Yehui and Xiong, Hao and Yang, Nianzu and Xiao, Tailong and Yan, Junchi},
  booktitle={The Twelfth International Conference on Learning Representations},
  year={2024}
}

@inproceedings{tangquadim,
  title={QuaDiM: A Conditional Diffusion Model For Quantum State Property Estimation},
  author={Tang, Yehui and Long, Mabiao and Yan, Junchi},
  booktitle={The Thirteenth International Conference on Learning Representations},
  year={2025}
}

@article{blume2010optimal,
  title={Optimal, reliable estimation of quantum states},
  author={Blume-Kohout, Robin},
  journal={New Journal of Physics},
  volume={12},
  number={4},
  pages={043034},
  year={2010},
  publisher={IOP Publishing}
}

@article{granade2016practical,
  title={Practical bayesian tomography},
  author={Granade, Christopher and Combes, Joshua and Cory, DG},
  journal={new Journal of Physics},
  volume={18},
  number={3},
  pages={033024},
  year={2016},
  publisher={IOP Publishing}
}

@article{lukens2020practical,
  title={A practical and efficient approach for Bayesian quantum state estimation},
  author={Lukens, Joseph M and Law, Kody JH and Jasra, Ajay and Lougovski, Pavel},
  journal={New Journal of Physics},
  volume={22},
  number={6},
  pages={063038},
  year={2020},
  publisher={IOP Publishing}
}

@article{lohani2023demonstration,
  title={Demonstration of machine-learning-enhanced Bayesian quantum state estimation},
  author={Lohani, Sanjaya and Lukens, Joseph M and Davis, Atiyya A and Khannejad, Amirali and Regmi, Sangita and Jones, Daniel E and Glasser, Ryan T and Searles, Thomas A and Kirby, Brian T},
  journal={New Journal of Physics},
  volume={25},
  number={8},
  pages={083009},
  year={2023},
  publisher={IOP Publishing}
}

@article{quek2021adaptive,
  title={Adaptive quantum state tomography with neural networks},
  author={Quek, Yihui and Fort, Stanislav and Ng, Hui Khoon},
  journal={npj Quantum Information},
  volume={7},
  number={1},
  pages={105},
  year={2021},
  publisher={Nature Publishing Group UK London}
}

@article{torlai2018neural,
  title={Neural-network quantum state tomography},
  author={Torlai, Giacomo and Mazzola, Guglielmo and Carrasquilla, Juan and Troyer, Matthias and Melko, Roger and Carleo, Giuseppe},
  journal={Nature physics},
  volume={14},
  number={5},
  pages={447--450},
  year={2018},
  publisher={Nature Publishing Group UK London}
}

@article{palmieri2024enhancing,
  title={Enhancing quantum state tomography via resource-efficient attention-based neural networks},
  author={Palmieri, Adriano Macarone and M{\"u}ller-Rigat, Guillem and Srivastava, Anubhav Kumar and Lewenstein, Maciej and Rajchel-Mieldzio{\'c}, Grzegorz and P{\l}odzie{\'n}, Marcin},
  journal={Physical Review Research},
  volume={6},
  number={3},
  pages={033248},
  year={2024},
  publisher={APS}
}

@article{zhong2022quantum,
  title={Quantum state tomography inspired by language modeling},
  author={Zhong, Lu and Guo, Chu and Wang, Xiaoting},
  journal={arXiv preprint arXiv:2212.04940},
  year={2022}
}

@article{ma2025tomography,
  title={Tomography of quantum states from structured measurements via quantum-aware transformer},
  author={Ma, Hailan and Sun, Zhenhong and Dong, Daoyi and Chen, Chunlin and Rabitz, Herschel},
  journal={IEEE Transactions on Cybernetics},
  year={2025},
  publisher={IEEE}
}

@article{huang2020predicting,
  title={Predicting many properties of a quantum system from very few measurements},
  author={Huang, Hsin-Yuan and Kueng, Richard and Preskill, John},
  journal={Nature Physics},
  volume={16},
  number={10},
  pages={1050--1057},
  year={2020},
  publisher={Nature Publishing Group UK London}
}

@book{nielsen2010quantum,
  title={Quantum computation and quantum information},
  author={Nielsen, Michael A and Chuang, Isaac L},
  year={2010},
  publisher={Cambridge university press}
}

@incollection{greenberger1989going,
  title={Going beyond Bell’s theorem},
  author={Greenberger, Daniel M and Horne, Michael A and Zeilinger, Anton},
  booktitle={Bell’s theorem, quantum theory and conceptions of the universe},
  pages={69--72},
  year={1989},
  publisher={Springer}
}

@article{greenberger1990bell,
  title={Bell’s theorem without inequalities},
  author={Greenberger, Daniel M and Horne, Michael A and Shimony, Abner and Zeilinger, Anton},
  journal={American Journal of Physics},
  volume={58},
  number={12},
  pages={1131--1143},
  year={1990},
  publisher={American Association of Physics Teachers}
}

@article{cha2025efficient,
  title={Efficient sampling for Pauli-measurement-based shadow tomography in direct fidelity estimation},
  author={Cha, Hyunho and Lee, Jungwoo},
  journal={Physical Review A},
  volume={112},
  number={3},
  pages={032427},
  year={2025},
  publisher={APS}
}

@article{zyczkowski2001induced,
  title={Induced measures in the space of mixed quantum states},
  author={Zyczkowski, Karol and Sommers, Hans-J{\"u}rgen},
  journal={Journal of Physics A: Mathematical and General},
  volume={34},
  number={35},
  pages={7111},
  year={2001},
  publisher={IOP Publishing}
}

@article{bravyi2019simulation,
  title={Simulation of quantum circuits by low-rank stabilizer decompositions},
  author={Bravyi, Sergey and Browne, Dan and Calpin, Padraic and Campbell, Earl and Gosset, David and Howard, Mark},
  journal={Quantum},
  volume={3},
  pages={181},
  year={2019},
  publisher={Verein zur F{\"o}rderung des Open Access Publizierens in den Quantenwissenschaften}
}

@article{aaronson2004improved,
  title={Improved simulation of stabilizer circuits},
  author={Aaronson, Scott and Gottesman, Daniel},
  journal={Physical Review A—Atomic, Molecular, and Optical Physics},
  volume={70},
  number={5},
  pages={052328},
  year={2004},
  publisher={APS}
}

@article{dehaene2003clifford,
  title={Clifford group, stabilizer states, and linear and quadratic operations over GF (2)},
  author={Dehaene, Jeroen and De Moor, Bart},
  journal={Physical Review A},
  volume={68},
  number={4},
  pages={042318},
  year={2003},
  publisher={APS}
}

@article{bengio2005curse,
  title={The curse of highly variable functions for local kernel machines},
  author={Bengio, Yoshua and Delalleau, Olivier and Roux, Nicolas},
  journal={Advances in neural information processing systems},
  volume={18},
  year={2005}
}

@inproceedings{telgarsky2016benefits,
  title={Benefits of depth in neural networks},
  author={Telgarsky, Matus},
  booktitle={Conference on learning theory},
  pages={1517--1539},
  year={2016},
  organization={PMLR}
}

@inproceedings{eldan2016power,
  title={The power of depth for feedforward neural networks},
  author={Eldan, Ronen and Shamir, Ohad},
  booktitle={Conference on learning theory},
  pages={907--940},
  year={2016},
  organization={PMLR}
}

@article{daniely2020learning,
  title={Learning parities with neural networks},
  author={Daniely, Amit and Malach, Eran},
  journal={Advances in Neural Information Processing Systems},
  volume={33},
  pages={20356--20365},
  year={2020}
}

@article{sugiyama2013precision,
  title={Precision-guaranteed quantum tomography},
  author={Sugiyama, Takanori and Turner, Peter S and Murao, Mio},
  journal={Physical Review Letters},
  volume={111},
  number={16},
  pages={160406},
  year={2013},
  publisher={APS}
}

@article{guctua2020fast,
  title={Fast state tomography with optimal error bounds},
  author={Gu{\c{t}}{\u{a}}, Madalin and Kahn, Jonas and Kueng, Richard and Tropp, Joel A},
  journal={Journal of Physics A: Mathematical and Theoretical},
  volume={53},
  number={20},
  pages={204001},
  year={2020},
  publisher={IOP Publishing}
}

@inproceedings{lee2019set,
  title={Set transformer: A framework for attention-based permutation-invariant neural networks},
  author={Lee, Juho and Lee, Yoonho and Kim, Jungtaek and Kosiorek, Adam and Choi, Seungjin and Teh, Yee Whye},
  booktitle={International conference on machine learning},
  pages={3744--3753},
  year={2019},
  organization={PMLR}
}

@article{brydges2019probing,
  title={Probing R{\'e}nyi entanglement entropy via randomized measurements},
  author={Brydges, Tiff and Elben, Andreas and Jurcevic, Petar and Vermersch, Beno{\^\i}t and Maier, Christine and Lanyon, Ben P and Zoller, Peter and Blatt, Rainer and Roos, Christian F},
  journal={Science},
  volume={364},
  number={6437},
  pages={260--263},
  year={2019},
  publisher={American Association for the Advancement of Science}
}

@article{hastings2010measuring,
  title={Measuring Renyi entanglement entropy in quantum Monte Carlo simulations},
  author={Hastings, Matthew B and Gonz{\'a}lez, Iv{\'a}n and Kallin, Ann B and Melko, Roger G},
  journal={Physical review letters},
  volume={104},
  number={15},
  pages={157201},
  year={2010},
  publisher={APS}
}

@article{islam2015measuring,
  title={Measuring entanglement entropy in a quantum many-body system},
  author={Islam, Rajibul and Ma, Ruichao and Preiss, Philipp M and Eric Tai, M and Lukin, Alexander and Rispoli, Matthew and Greiner, Markus},
  journal={Nature},
  volume={528},
  number={7580},
  pages={77--83},
  year={2015},
  publisher={Nature Publishing Group UK London}
}

@article{chen2021robust,
  title={Robust shadow estimation},
  author={Chen, Senrui and Yu, Wenjun and Zeng, Pei and Flammia, Steven T},
  journal={PRX Quantum},
  volume={2},
  number={3},
  pages={030348},
  year={2021},
  publisher={APS}
}

@article{lange2023adaptive,
  title={Adaptive quantum state tomography with active learning},
  author={Lange, Hannah and Kebri{\v{c}}, Matja{\v{z}} and Buser, Maximilian and Schollw{\"o}ck, Ulrich and Grusdt, Fabian and Bohrdt, Annabelle},
  journal={Quantum},
  volume={7},
  pages={1129},
  year={2023},
  publisher={Verein zur F{\"o}rderung des Open Access Publizierens in den Quantenwissenschaften}
}

@inproceedings{aaronson2018shadow,
  title={Shadow tomography of quantum states},
  author={Aaronson, Scott},
  booktitle={Proceedings of the 50th annual ACM SIGACT symposium on theory of computing},
  pages={325--338},
  year={2018}
}

@article{degen2017quantum,
  title={Quantum sensing},
  author={Degen, Christian L and Reinhard, Friedemann and Cappellaro, Paola},
  journal={Reviews of modern physics},
  volume={89},
  number={3},
  pages={035002},
  year={2017},
  publisher={APS}
}

@article{montanaro2016quantum,
  title={Quantum algorithms: an overview},
  author={Montanaro, Ashley},
  journal={npj Quantum Information},
  volume={2},
  number={1},
  pages={1--8},
  year={2016},
  publisher={Nature Publishing Group}
}

@article{cozzolino2019high,
  title={High-dimensional quantum communication: benefits, progress, and future challenges},
  author={Cozzolino, Daniele and Da Lio, Beatrice and Bacco, Davide and Oxenl{\o}we, Leif Katsuo},
  journal={Advanced Quantum Technologies},
  volume={2},
  number={12},
  pages={1900038},
  year={2019},
  publisher={Wiley Online Library}
}

@inproceedings{haah2016sample,
  title={Sample-optimal tomography of quantum states},
  author={Haah, Jeongwan and Harrow, Aram W and Ji, Zhengfeng and Wu, Xiaodi and Yu, Nengkun},
  booktitle={Proceedings of the forty-eighth annual ACM symposium on Theory of Computing},
  pages={913--925},
  year={2016}
}

@inproceedings{o2016efficient,
  title={Efficient quantum tomography},
  author={O'Donnell, Ryan and Wright, John},
  booktitle={Proceedings of the forty-eighth annual ACM symposium on Theory of Computing},
  pages={899--912},
  year={2016}
}

@article{gross2010quantum,
  title={Quantum state tomography via compressed sensing},
  author={Gross, David and Liu, Yi-Kai and Flammia, Steven T and Becker, Stephen and Eisert, Jens},
  journal={Physical review letters},
  volume={105},
  number={15},
  pages={150401},
  year={2010},
  publisher={APS}
}

@article{buvzek1998reconstruction,
  title={Reconstruction of quantum states of spin systems: From quantum Bayesian inference to quantum tomography},
  author={Bu{\v{z}}ek, V and Derka, Radoslav and Adam, G and Knight, Peter L},
  journal={Annals of Physics},
  volume={266},
  number={2},
  pages={454--496},
  year={1998},
  publisher={Elsevier}
}

@article{schack2001quantum,
  title={Quantum bayes rule},
  author={Schack, Ruediger and Brun, Todd A and Caves, Carlton M},
  journal={Physical Review A},
  volume={64},
  number={1},
  pages={014305},
  year={2001},
  publisher={APS}
}

@article{wei2024neural,
  title={Neural-shadow quantum state tomography},
  author={Wei, Victor and Coish, WA and Ronagh, Pooya and Muschik, Christine A},
  journal={Physical Review Research},
  volume={6},
  number={2},
  pages={023250},
  year={2024},
  publisher={APS}
}

@inproceedings{tang2024ssl4q,
  title={SSL4Q: semi-supervised learning of quantum data with application to quantum state classification},
  author={Tang, Yehui and Yang, Nianzu and Long, Mabiao and Yan, Junchi},
  booktitle={Forty-first International Conference on Machine Learning},
  year={2024}
}

@article{koh2022classical,
  title={Classical shadows with noise},
  author={Koh, Dax Enshan and Grewal, Sabee},
  journal={Quantum},
  volume={6},
  pages={776},
  year={2022},
  publisher={Verein zur F{\"o}rderung des Open Access Publizierens in den Quantenwissenschaften}
}

@article{wu2024error,
  title={Error-mitigated fermionic classical shadows on noisy quantum devices},
  author={Wu, Bujiao and Koh, Dax Enshan},
  journal={npj Quantum Information},
  volume={10},
  number={1},
  pages={39},
  year={2024},
  publisher={Nature Publishing Group UK London}
}
\bibliographystyle{plainnat}

\clearpage

\appendix

\section{Quantum states and density matrices}
\label{sec:quantum_state_density_matrix_intro}

A quantum state of an $n$-qubit system with Hilbert space $\mathcal H\cong\mathbb C^{2^n}$ may be either \emph{pure} or \emph{mixed}. A \emph{pure state} is represented (up to global phase) by a unit vector $\ket\psi\in\mathcal H$.  Equivalently, one may describe it by the rank-one density operator \(\rho = \ketbra\psi \). A \emph{mixed state} arises from a statistical ensemble $\{(p_i,\ket{\psi_i})\}$ of pure states, with $p_i\ge0$ and $\sum_i p_i=1$.  Its density operator is \(\sum_i p_i \ketbra{\psi_i}\). In either case, the \emph{density matrix} $\rho$ satisfies \(\rho \succeq 0\) and \(\tr(\rho)=1\), i.e., it is positive semi-definite with unit trace. The expectation value of any observable $O\in\mathcal O(\mathcal H)$ is given by \(\langle O\rangle = \tr(\rho O)\). Finally, if the system is bipartitioned as $AB$, the state of subsystem $A$ alone is described by the \emph{reduced density matrix} \(\rho_A = \tr_B\left(\rho_{AB}\right)\), where $\tr_B$ denotes the partial trace over subsystem $B$.

\section{Classical shadows}
\label{sec:classical_shadow_intro}

In the classical shadows protocol \citep{huang2020predicting}, one performs randomized measurements on many independent copies of an unknown state $\rho$. Each measurement outcome is used to construct a \emph{snapshot} $\hat{\rho}$---an unbiased estimator of the state---such that by averaging these snapshots one can estimate any linear function of $\rho$. For example, for a Hermitian observable $O$, the classical shadows estimator satisfies $\E[\tr(\hat{\rho} O)] = \tr(\rho O)$, assuming $\E[\hat{\rho}] = \rho$ by construction.

\begin{algorithm}
\caption{Classical shadow estimation of \(\tr(\rho O)\) \citep{huang2020predicting}}
\label{alg:shadow_estimation}
\KwIn{Unknown state \( \rho \in \mathcal{S}(\mathcal{H}) \), number of copies \(N \in \N\), observable \(O \in \mathcal{O}(\mathcal{H})\), group \(\G \subset \text{U}(2^n)\).}
\KwOut{Estimate of \( \tr ( \rho O ) \).}
Initialize the shadow protocol with \(\G\) and the corresponding channel \(\mathcal{E}_\G\).\\
\( \texttt{sum} \leftarrow 0 \)\\
\For{$k = 1$ \KwTo $N$}{
    Uniformly sample \(U\) from \(\G\).\\
    Measure \(\rho\) in the \(U\) basis and get \(\ket{\hat{\mathbf{b}}}\).\\
    \(\texttt{sum}\,\texttt{+=}\,\tr\left(\mathcal{E}_\G^{-1}(U^\dagger\ket{\hat{\mathbf{b}}}\bra{\hat{\mathbf{b}}}U)O\right)\)
}
\Return{\(\textnormal{\texttt{sum}} / N\)}
\end{algorithm}

To illustrate, consider the fidelity between quantum states---a widely used measure of similarity. Given two density matrices \(\rho\) and \(\sigma\), the fidelity is commonly defined as
\begin{equation*}
\mathcal{F}(\rho,\sigma) = \left(\tr\sqrt{\sqrt{\rho}\sigma\sqrt{\rho}}\right)^2.
\end{equation*}
In the special case where the second state is a pure state \(\ket{\psi}\), we have
\begin{align*}
\mathcal{F}(\rho,\ketbra{\psi})=\bra{\psi}\rho\ket{\psi} = \tr (\rho O),
\end{align*}
where \(O = \ketbra{\psi}\).

In the most general formulation (see Algorithm~\ref{alg:shadow_estimation}), one fixes a finite unitary ensemble $\G\subseteq U(2^n)$, samples \(U\sim\mathrm{Unif}(\G)\), measures $U\rho U^\dagger$ in the computational basis, and then applies the corresponding inverse channel $\mathcal{E}_{\G}^{-1}$ to form an unbiased snapshot of $\rho$.

Typically, Algorithm~\ref{alg:shadow_estimation} is instantiated with one of two standard ensembles \(\G\). The first kind is \emph{random Pauli measurements}, which sample from local Pauli $\{X, Y, Z\}$ measurements uniformly randomly for each qubit. The second kind is \emph{random Clifford measurements}, which sample the rotation $U \rho U^\dagger$ from the $n$-qubit Clifford group $\mathrm{Cl}(2^n)$. In either case, the computational basis measurement outcome can be represented as an $n$-bit string $\hat{b} \in \{0,1\}^n$ and the corresponding ket $\ket{\hat{b}} \in \mathcal{H}$. It is well known that, given access to repeated measurements, one can reconstruct the density matrix $\rho$ of an unknown quantum state.

For random Pauli measurements, the corresponding snapshot is given by
\begin{equation}
\label{equation:pauli_snapshot}
\hat{\rho} = \bigotimes_{i=1}^n \hat{\rho}_i = \bigotimes_{i=1}^n \left( 3 U_i^\dagger \ketbra{\hat{b}_i} U_i - \I \right),
\end{equation}
where each $U_i$ is independently sampled uniformly at random from the set $\{\mathbb{I}, H, HS^\dagger\}$. This corresponds to performing local measurements in a randomly chosen basis from $\{X, Y, Z\}$. As will be discussed in the main text, this work considers DFE with Pauli measurements, where the target state is the GHZ state \citep{greenberger1989going, greenberger1990bell}. In such cases, more efficient alternatives to the direct use of~\eqref{equation:pauli_snapshot} are available, and we adopt one of these optimized methods.

For random Clifford measurements, the corresponding snapshot is given by
\begin{equation}
\label{equation:clifford_snapshot}
\hat{\rho} = \left(2^n + 1 \right) U^\dagger \ket{\hat{\mathbf{b}}}\bra{\hat{\mathbf{b}}} U - \I^{\otimes n},
\end{equation}
where $U$ is sampled uniformly from the $n$-qubit Clifford group $\mathrm{Cl}(2^n)$ \citep{nielsen2010quantum}.

The function estimator effectively computes the sample mean of $\tr(\hat{\rho} O)$ over the snapshots, treating each outcome independently. While this provides an unbiased estimate for linear functionals, it does not incorporate prior knowledge about $\rho$ and suffers from high variance because each snapshot $\hat{\rho}$ is generally not a valid quantum state. In contrast, a Bayesian approach would update a prior distribution for $\rho$ based on the measurement data and could potentially yield improved estimates. In this work, we focus on applying Bayesian inference to scalar functions $f(\rho)$ of the state (such as expectation values of observables or entanglement metrics), using classical shadow data as input.

\section{Proof of Observation~\ref{observation:permutation_invariance}}
\label{sec:permutation_invariance_proof}
\begin{proof}
This is a direct consequence of the likelihood principle. Let \(\mathbb{G}\subset\text{U}(2^n)\) be a finite group. Let's say \(\mathbf{v}^{(k)}\) encodes measurement basis \(U_{i^{(k)}}\in\mathbb{G}\) and \(n\)-bit outcome \(\hat{\mathbf{b}}^{(k)}\). Define \(\mathcal{P}\in\R^{|\mathbb{G}|\times2^n}\) such that
\begin{equation}
\label{equation:measurement_matrix}
\mathcal{P}_{i,j} := \langle j-1|U_i\rho U_i^\dagger|j-1\rangle,
\end{equation}
where \(U_i\in\mathbb{G}\). If the measurement is informationally complete, then \(\mathcal{P}\) completely determines \(\rho\), and vice versa. So estimating the posterior of \(\rho\) amounts to estimating the posterior distribution of \(\mathcal{P}\):
\[
\pi(\mathcal{P} \mid \mathcal{D}_N),
\]
where
\[
\mathcal{D}_k := \left(\mathbf{v}^{(1)}, \dots , \mathbf{v}^{(k)}\right).
\]
But the likelihood is
\begin{align*}
p_\mathcal{P}(\mathcal{D}_N) & = \prod_{k=1}^N P\left(\mathbf{v}^{(k)} \mid \mathcal{D}_{k-1}\right)\\
& = \prod_{k=1}^N P\left(i^{(k)}, \hat{\mathbf{b}}^{(k)} \mid \mathcal{D}_{k-1}\right)\\
& = \prod_{k=1}^N P\left(i^{(k)} \mid \mathcal{D}_{k-1}\right) \mathcal{P}_{i^{(k)}, \hat{\mathbf{b}}^{(k)}}\\
& = \left[\prod_{k=1}^N P\left(i^{(k)} \mid \mathcal{D}_{k-1}\right)\right] \left[\prod_{k=1}^N \mathcal{P}_{i^{(k)}, \hat{\mathbf{b}}^{(k)}}\right],
\end{align*}
So
\[
\pi(\mathcal{P}\mid\mathcal{D}_N) \propto \pi(\mathcal{P}) \prod_{k=1}^N \mathcal{P}_{i^{(k)}, \hat{\mathbf{b}}^{(k)}} = \pi(\mathcal{P}) \prod_{i=1}^{|\mathbb{G}|} \prod_{j=1}^{2^n} \mathcal{P}_{i,j}^{e_{i,j}},
\]
where
\[
e_{i,j} := \left|\left\{k | i^{(k)} = i, \hat{\mathbf{b}}^{(k)} = j\right\}\right|.
\]
Since any permutation of the data \(\mathcal{D}_N\) leaves all \(e_{i,j}\) unchanged, it leaves the posterior unchanged.
\end{proof}

\section{Set transformer architecture}
\label{sec:set_transformer_diagram}

\begin{figure}[h]
    \centering
    \includegraphics[width=\linewidth]{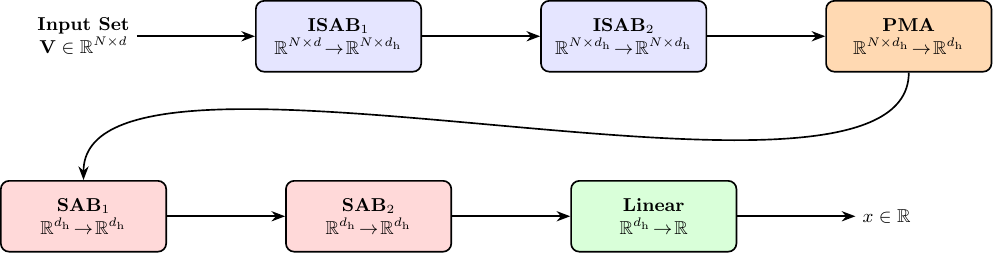}
    \caption{\justifying The set transformer architecture employed for Bayesian estimation of functions of quantum states.}
    \label{fig:Set_Transformer_Architecture}
\end{figure}

\section{Proof of Observation~\ref{observation:bit_flip_noise}}
\label{sec:bit_flip_noise_proof}
\begin{proof}
Define the matrix \(\mathcal{P}\) as introduced in \eqref{equation:measurement_matrix}. Also, define \(\mathcal{P}^{(\lambda)}\in\R^{|\mathbb{G}|\times2^n}\) such that \(\mathcal{P}^{(\lambda)}_{i,j}\) denotes the probability of measuring \(|j-1\rangle\) from \(U_i\rho U_i^\dagger\) under bit-flip noise with parameter \(\lambda\), where \(U_i\in\mathbb{G}\). Then \(\mathcal{P}^{(\lambda)}\) satisfies
\[
\left(\mathcal{P}_{i,:}^{(\lambda)}\right)^T = M (\mathcal{P}_{i,:})^T, \quad \text{where } M = \begin{pmatrix}
1-\lambda & \lambda\\
\lambda & 1 - \lambda
\end{pmatrix}^{\otimes n}.
\]
If \(\lambda\ne\frac{1}{2}\), then \(M\) is invertible, and thus \(\mathcal{P}\) can be recovered from \(\mathcal{P}^{(\lambda)}\).
\end{proof}

\clearpage

\section{Encoding of measurement data}
\label{sec:encoding_diagram}

\begin{figure}[h]
    \centering
    \begin{subfigure}[b]{0.44\linewidth}
        \centering
        \includegraphics[page=1, width=\linewidth]{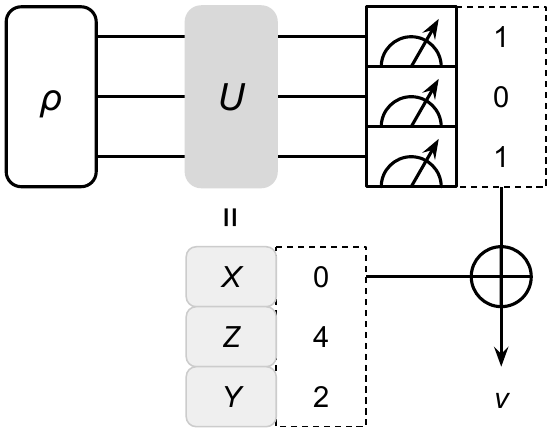}
        \caption{}
        \label{fig:Pauli_encoding}
    \end{subfigure}
    \hfill
    \begin{subfigure}[b]{\linewidth}
        \centering
        \includegraphics[page=2, width=\linewidth]{figures/diagrams.pdf}
        \caption{}
        \label{fig:Clifford_encoding}
    \end{subfigure}
    \caption{\justifying Encoding of quantum measurement data for Bayesian conditioning: (a) random Pauli measurements and (b) random Clifford measurements.}
    \label{fig:Pauli_Clifford_encoding}
\end{figure}

\clearpage

\section{Deferred algorithmic details}
\label{sec:deferred_algorithms}

\begin{algorithm}
\caption{DFE for the GHZ state with random Pauli measurements \citep{cha2025efficient}}
\label{alg:ghz_dfe}
\KwIn{Unknown state \( \rho \in \mathcal{S}(\mathcal{H}) \), number of copies \(N \in \N\).}
\KwOut{Estimate of \( f(\rho) = \tr ( \rho \ketbra{\psi_{\text{GHZ}}} ) \).}
\( \texttt{sum} \leftarrow 0 \)\\
\For{$k = 1$ \KwTo $N$}{
    Sample \(X \sim \mathcal{U}_{(0,1)}\).\\
    \If{\(X<1/3\)}{
        Measure \(\rho\) in the computational basis and get \(\ket{\hat{\mathbf{b}}}\).\\
        \( F = 3\left(\delta_{\hat{\mathbf{b}}, \mathbf{0}} + \delta_{\hat{\mathbf{b}}, \mathbf{1}}\right) / 2 - 3 / 4 \)\label{line:dfe_ghz_first_branch_sample}\\
        \(\texttt{sum}\,\texttt{+=}\,F\)
    } \Else {
        Sample $P_i \in \{X,Y\}$ uniformly and independently for $1 \leq i \leq n - 1$.\\
        Choose $P_n \in \{X,Y\}$ such that $|\{1 \le i \le n \mid P_i = Y\}| \mod{2} = 0$.\\
        Measure \(\rho\) in the Pauli basis \( P = (P_1, \ldots , P_n) \) and get \(\ket{\hat{\mathbf{b}}}\).\\
        \( F = 3 \cdot (-1)^{|\{i \mid P_i = Y\}| / 2 + |\{i \mid \hat{b}_i = 1\}|} / 4 \)\label{line:dfe_ghz_second_branch_sample}\\
        \(\texttt{sum}\,\texttt{+=}\,F\)
    }
}
\Return{\( \hat{F} = \textnormal{\texttt{sum}} / N + 1/4 \)}
\end{algorithm}

\clearpage

\begin{algorithm}
\caption{Second-order R\'enyi entanglement entropy estimation \citep{huang2020predicting}}
\label{alg:renyi}
\KwIn{Unknown state \( \rho \in \mathcal{S}(\mathcal{H}) \), number of copies \(N \in \N\), subsystem index set \(A\), group \(\G \subset \text{U}(2^n)\).}
\KwOut{Estimate of \( \tr ( (\rho \otimes \rho) S_A ) \).}
Initialize the shadow protocol with \(\G\) and the corresponding channel \(\mathcal{E}_\G\).\\
\For{$k = 1$ \KwTo $N$}{
Uniformly sample \(U\) from \(\G\).\\
Measure \(\rho\) in the \(U\) basis and get \(\ket{\hat{\mathbf{b}}}\).\\
\(\hat{\rho}^{(k)}\leftarrow\mathcal{E}_\G^{-1}(U^\dagger\ket{\hat{\mathbf{b}}}\bra{\hat{\mathbf{b}}}U)\)
}
\(F\leftarrow\frac{1}{N(N-1)}\sum_{k\ne l}\tr((\hat{\rho}^{(k)}\otimes\hat{\rho}^{(l)})S_A)\)\label{line:renyi_snapshot}\\
\Return{F}
\end{algorithm}

\section{Proof of Observation~\ref{observation:dfe_estimator_variance}}
\label{sec:dfe_estimator_variance_proof}

\begin{proof}
Note that
\[
\E[F] = \E[\hat{F}]-\frac{1}{4} = f(\rho)-\frac{1}{4}
\]
and \(F\) takes values in \(\{3/4, -3/4\}\). Therefore, \(f(\rho)\) determines \(\text{Var}(F)\) since
\[
f(\rho)-\frac{1}{4} = \frac{3}{4}(2\cdot P(F=3/4)-1).
\]
This gives \eqref{equation:var_F}.
\end{proof}

\section{Detailed restatement and proof of Lemma~\ref{lemma:algorithm_locally_optimal}}
\label{sec:algorithm_locally_optimal_proof}

We first define the class of adaptive strategies that sample from the same set of measurement settings as Algorithm~\ref{alg:ghz_dfe} with reweighted estimators, where we take the \emph{default policy} \(p(P)\) from Algorithm~\ref{alg:ghz_dfe}. The \emph{adaptive policy} is denoted as \(q_t(P \mid \mathcal{H}_{t-1})\), where
\[
\mathcal{H}_{t-1}=\left( \left( P^{(0)},\hat{\mathbf{b}}^{(0)} \right), \dots , \left( P^{(t-1)},\hat{\mathbf{b}}^{(t-1)} \right) \right)
\]
is the history of measurement outcomes. Define
\[
F^{(t)} = F\left(P^{(t)},\hat{\mathbf{b}}^{(t)}\right),
\]
which denotes the sample in either line \ref{line:dfe_ghz_first_branch_sample} or \ref{line:dfe_ghz_second_branch_sample} in Algorithm~\ref{alg:ghz_dfe}. Then the reweighted (unbiased) estimator is naturally defined as
\[
\hat{F} = \frac{1}{N} \sum_{t=1}^N w_t F^{(t)}+\frac{1}{4},
\]
where
\[
w_t = \frac{p\left(P^{(t)}\right)}{q_t\left(P^{(t)} \mid \mathcal{H}_{t-1}\right)}.
\]
Write
\begin{equation}
\label{equation:f_hat_alternate_expression}
\hat{F} = \frac{1}{N} \sum_{t=1}^N (\mu+D_t)+\frac{1}{4},
\end{equation}
where \(\mu:=f(\rho)-1/4\) and \(D_t:=w_tF^{(t)}-\mu\). Then we have \(\E[D_t \mid \mathcal{H}_{t-1}] = 0\), so \(\{D_t, \mathcal{H}_t\}\) is a \emph{martingale difference sequence} and the following cross terms vanish:
\begin{equation}
\label{equation:martingale_cross_vanish}
\E[D_t D_{t^\prime}] = \E[D_t \E[D_{t^\prime} \mid \mathcal{H}_{t^\prime-1}]] = 0 \quad (t<t^\prime).
\end{equation}
Meanwhile, from \eqref{equation:f_hat_alternate_expression} we have
\[
\text{Var}(\hat{F}) = \frac{1}{N^2} \text{Var}\left( \sum_{t=1}^N D_t \right).
\]
But from \eqref{equation:martingale_cross_vanish},
\[
\text{Var}\left( \sum_{t=1}^N D_t \right) = \sum_{t=1}^N \E[D_t^2] = \sum_{t=1}^N \E [ \E[D_t^2 \mid \mathcal{H}_{t-1}] ] = \sum_{t=1}^N \E [ \text{Var}(w_tF^{(t)} \mid \mathcal{H}_{t-1}) ].
\]
But
\begin{equation}
\label{equation:objective_each_t}
\text{Var}(w_tF^{(t)} \mid \mathcal{H}_{t-1}) = \sum_{P^{(t)}} \frac{p\left(P^{(t)}\right)^2}{q_t\left(P^{(t)} \mid \mathcal{H}_{t-1}\right)} \E_{\hat{\textbf{b}}^{(t)}}[F^{(t)2} \mid P^{(t)}] - \mu^2.
\end{equation}
The variance \eqref{equation:objective_each_t} is to be minimized. From Algorithm~\ref{alg:ghz_dfe}, we see that \(|F|=3/4\) regardless of \(t\) and the measurement outcome. Therefore, for each \(t\), the problem reduces to the following minimization:
\[
\underset{q_t}{\min} \sum_{P^{(t)}} \frac{p\left(P^{(t)}\right)^2}{q_t\left(P^{(t)} \mid \mathcal{H}_{t-1}\right)} \quad \text{subject to} \quad \sum_{P^{(t)}} q_t\left(P^{(t)} \mid \mathcal{H}_{t-1}\right) = 1.
\]
From the Cauchy–Schwarz inequality, the optimal policy is always \(q_t \propto p\), i.e., \(q_t = p\).\qed

\section{Proof of Observation~\ref{observarion:renyi_pauli_snapshot_is_local}}
\label{sec:renyi_pauli_snapshot_is_local_proof}

\begin{proof}
From \eqref{equation:pauli_snapshot}, the snapshot \(\hat{\rho}\) factors as
\[
\hat{\rho} = \hat{\rho}_A \otimes \hat{\rho}_B.
\]
Therefore,
\begin{align*}
\tr\left(\left(\hat{\rho}^{(k)}\otimes\hat{\rho}^{(l)}\right)S_A\right) & = \tr\left(\left(\hat{\rho}_A^{(k)}\otimes\hat{\rho}_B^{(k)}\otimes\hat{\rho}_A^{(l)}\otimes\hat{\rho}_B^{(l)}\right)S_A\right) \\
& = \tr\left(\text{SWAP}_{A,A}\left(\hat{\rho}_A^{(k)}\otimes\hat{\rho}_A^{(l)}\right)\right) \tr\left(\hat{\rho}_B^{(k)}\otimes\hat{\rho}_B^{(l)}\right)\\
& = \tr\left(\hat{\rho}_A^{(k)}\hat{\rho}_A^{(l)}\right) \tr\left(\hat{\rho}_B^{(k)}\right) \tr\left(\hat{\rho}_B^{(l)}\right)\\
& = \tr\left(\hat{\rho}_A^{(k)}\hat{\rho}_A^{(l)}\right),
\end{align*}
where the last equality follows from the fact that each multiplicand in \eqref{equation:pauli_snapshot} has unit trace.
\end{proof}

\begin{remark}
From Observation~\ref{observarion:renyi_pauli_snapshot_is_local}, line~\ref{line:renyi_snapshot} in Algorithm~\ref{alg:renyi} for random Pauli measurements can be calculated efficiently as
\[
\tr\left(\hat{\rho}_A^{(k)}\hat{\rho}_A^{(l)}\right) = \prod_{i \in A} \tr\left(\hat{\rho}_i^{(k)}\hat{\rho}_i^{(l)}\right),
\]
where
\begin{equation}
\label{equation:local_renyi_snapshot}
\tr\left(\hat{\rho}_i^{(k)}\hat{\rho}_i^{(l)}\right) = \begin{cases}
5 & U_i^{(k)} = U_i^{(l)} \text{ and } \hat{b}_i^{(k)} = \hat{b}_i^{(l)} \\
-4 & U_i^{(k)} = U_i^{(l)} \text{ and } \hat{b}_i^{(k)} \ne \hat{b}_i^{(l)} \\
1/2 & U_i^{(k)} \ne U_i^{(l)} \\
\end{cases}.
\end{equation}
Also, for random Clifford measurements, line~\ref{line:renyi_snapshot} in Algorithm~\ref{alg:renyi} can be calculated in \(\text{poly}(n)\) time due to the Gottesman-Knill theorem and standard stabilizer-tableau update methods \citep{aaronson2004improved, nielsen2010quantum}.
\end{remark}


\end{document}